\documentclass[iop]{emulateapj}

\usepackage{nicefrac}
\usepackage[dvipsnames]{xcolor}
\usepackage{amsmath}
\usepackage{listings}
\usepackage{CJKutf8}
\usepackage{txfonts}

\makeatletter
\usepackage[unicode,breaklinks]{hyperref}
\usepackage[all]{hypcap}
\hypersetup{backref,
            bookmarks,
            dvips,
            pdfdisplaydoctitle,
            colorlinks=true,
            citecolor=blue,
            unicode}

\usepackage{etoolbox}
\patchcmd{\NAT@citex}
  {\@citea\NAT@hyper@{%
     \NAT@nmfmt{\NAT@nm}%
     \hyper@natlinkbreak{\NAT@aysep\NAT@spacechar}{\@citeb\@extra@b@citeb}%
     \NAT@date}}
  {\@citea\NAT@nmfmt{\NAT@nm}%
   \NAT@aysep\NAT@spacechar\NAT@hyper@{\NAT@date}}{}{}

\patchcmd{\NAT@citex}
  {\@citea\NAT@hyper@{%
     \NAT@nmfmt{\NAT@nm}%
     \hyper@natlinkbreak{\NAT@spacechar\NAT@@open\if*#1*\else#1\NAT@spacechar\fi}%
       {\@citeb\@extra@b@citeb}%
     \NAT@date}}
  {\@citea\NAT@nmfmt{\NAT@nm}%
   \NAT@spacechar\NAT@@open\if*#1*\else#1\NAT@spacechar\fi\NAT@hyper@{\NAT@date}}
  {}{}
\makeatother

\shorttitle{Block time step storage scheme}
\shortauthors{Cai et al.}

\begin{document}

\title{Block Time Step Storage Scheme for Astrophysical \textit{N}-body Simulations}

\author{Maxwell Xu Cai (\begin{CJK*}{UTF8}{gbsn}蔡栩\end{CJK*}\!)\altaffilmark{1,2}, Yohai Meiron (\begin{CJK*}{UTF8}{gbsn}林友海\end{CJK*}\!)\altaffilmark{2,1}, M.B.N. Kouwenhoven (\begin{CJK*}{UTF8}{gbsn}柯文采\end{CJK*}\!)\altaffilmark{2},\\
 Paulina Assmann\altaffilmark{3,1,4}, Rainer Spurzem\altaffilmark{1,2,5}}

\email{maxwell@nao.cas.cn}

\affil{
   \altaffilmark{1} National Astronomical Observatories, Chinese Academy of Sciences, 20A Datun Road, Chaoyang District, Beijing 100012, China\\
   \altaffilmark{2} Kavli Institute for Astronomy and Astrophysics, Peking University, 5 Yiheyuan Road, Haidian District, Beijing 100871, China\\
   \altaffilmark{3} Departamento de Astronom\'ia, Universidad de Chile, Camino El observatorio 1515, Las Condes, Santiago, Chile\\
   \altaffilmark{4} Departamento de Astronom\'ia, Universidad de Conc\'epcion, Casilla 160-C, Conc\'epcion, Chile\\
   \altaffilmark{5} Astronomisches Rechen-Institut, Zentrum f\"ur Astronomie, University of Heidelberg, M\"onchhofstrasse 12-14, 69120 Heidelberg, Germany\\
}

\submitted{Received 2014 December 9; accepted 2015 June 23}

\journalinfo{Accepted for publication in The Astrophysical Journal Supplement Series}

\begin{abstract}
Astrophysical research in recent decades has made significant progress thanks to the availability of various $N$-body simulation techniques. With the rapid development of high-performance computing technologies, modern simulations have been able to take the computing power of massively parallel clusters with more than $10^5$ GPU cores. While unprecedented accuracy and dynamical scales have been achieved, the enormous amount of data being generated continuously poses great challenges for the subsequent procedures of data analysis and archiving. As an urgent response to these challenges, in this paper we propose an adaptive storage scheme for simulation data, inspired by the block time step integration scheme found in a number of direct $N$-body integrators available nowadays. The proposed scheme, namely the block time step storage scheme, works by minimizing the data redundancy with assignments of data with individual output frequencies as required by the researcher. As demonstrated by benchmarks, the proposed scheme is applicable to a wide variety of simulations. Despite the main focus of developing a solution for direct $N$-body simulation data, the methodology is transferable for grid-based or tree-based simulations where hierarchical time stepping is used.
\end{abstract}


\section{Introduction}
The gravitational $N$-body problem has posed a challenge ever since it was mathematically formulated in the 17\textsuperscript{th} century by Isaac Newton. This problem of determining from initial conditions the future motion of $N$-bodies interacting gravitationally amongst themselves continues to be relevant in modern day astronomy and is investigated in the context of planetary systems, star clusters, galaxies and the Universe. Mathematically, it is posed as $3N$ coupled nonlinear second-order ordinary differential equations. The solution consists of the phase-space paths of all particles as functions of time, which generally cannot be expressed by algebraic expressions or integrals.

Gravitational $N$-body simulations are currently the preferred approach for finding these solutions. They use particles to represent gravitating objects and propagate the initial conditions in time by calculating the force acting on each particle, and advancing it in time in small steps.

With advances in technology, simulations have become elaborate enough to take full advantage of computing capabilities. Especially the availability of highly-parallelized computing facilities, some using hardware accelerators (such as \emph{GRAPE} \citep{makino98}\footnote{GRAPE: GRAvity piPEline}, \emph{FPGA}\footnote{FPGA: Field Programmable Gate Array} boards \citep{berczik07}, and more recently \emph{GPU}s\footnote{GPU: Graphics Processing Unit}) have contributed to recent progress in the field. Simulations are carried out with more particles than ever before, and for longer integration times. Modern simulations are also characterized by the inclusion of more detailed physical processes and requirements for higher numerical accuracy.

While modern powerful hardware has brought astrophysical simulations to unprecedented accuracy, complexities come with the problem of storing the results, which is done by writing to the hard disk some or all properties of some or all particles at some pre-specified times; this is often called ``taking a snapshot''. The snapshot files can later be processed to learn about the evolution of the system. The storage requirement is determined by four factors: (1) the number of particles, (2) the size of the data record per particle, (3) the output frequency and (4) total integration time. In order to capture the detailed physical processes, high time-resolution of output often necessary. There is thus a tradeoff between time-resolution and output size (or the availability of storage space and post-processing capabilities).

To illustrate this, consider some large cosmological simulations from the previous decade. The Millennium Simulation \citep{springel05} followed about $10^{10}$ particles for nearly a Hubble time, and produced only 64 snapshots of about 300 GB each. Similarly, the MareNostrum simulation \citep{stefan06} used $2\times10^9$ particles and saved 135 snapshots of 64 GB each (running for a similar physical time). Even almost a decade later, this data volume still poses a challenge for storage, and more crucially for transport over a network and for analysis. Thus, there is a gap between computing power and data processing and management capabilities.

Direct summation techniques are often preferred when studying a system in which accurate orbital integration is needed and encounters are important, and/or physical assumptions have to be minimized, such as globular clusters and planetary systems. In cosmological simulations, where the primary interest is to study the evolution of the large scale structure, one often uses Tree methods \citep{barnes86} or Fast Multipole Method \citep{greengard87}, which are generally much faster (for large $N$) but introduce an approximation to the force contributions from very distant particles. Direct $N$-body simulations thus generally use a much smaller number of particles, nowadays rarely more than $N = 10^6$.

Big data management has so far been in the domain of cosmological collisionless simulations due to the large number of particles. Despite the relatively small number of particles in direct $N$-body simulations, data output can be a challenge for this kind of simulations as well. The key challenge is that they attempt to follow accurately phenomena which happen on vastly different timescales: from white dwarf binaries which have orbital periods of less than one hour \citep{brown11} to stars orbiting in the outskirts of the cluster, which can take millions of years to complete one orbit. Calculating the evolution of the entire cluster based on the smallest time step or timescale is completely impractical (see an estimation in Section~\ref{sec:bts}), so most productive codes employ individual or hierarchical time step schemes like the Hermite scheme \citep{aarseth03}. Saving the output, however, is usually done using snapshots in much the same way as for the large cosmological simulations.

The main problem with the snapshot approach is that no information is stored about what happens between snapshots. Interpolating will not always yield useful information if the process of interest occurs on a much shorter timescale than the snapshot interval. Examples of this are close encounters that may create hypervelocity stars (e.g. \citealt{yu03}), resonances such as Kozai oscillations (e.g. \citealt{katz11}), evolution of planetary systems (e.g. \citealt{Hao13}) or supernova explosions. Those phenomena may be captured by the program and recorded separately, but there is no standardized way of doing so. For the same reason, it is difficult to make a smooth visualization of an energetically active subsystem. On the other hand, there might be redundant data for the dynamically inactive particles. For this reason, \cite{Farr12} proposed an adaptive approach of data output such that only recently changed data during the last output interval will be written to files. They also proposed the Particle Stream Data Format (PSDF), a YAML (Yet another Markup Language)\footnote{\tt http://www.yaml.org/} based structured text format to ensure machine-independence and flexibility for most simulation data.

Traditionally, snapshot files are simple \textsc{ascii} files containing a table where rows represent the particles and the columns represent their properties; a header may have some additional information such as the snapshot time. This format is used, for example, by the {\tt phiGRAPE} code \citep{harfst07}. While this scheme has some advantages being easy to process, human-readable, and machine-independent, it is not native to the machine representation of data and usually requires auxiliary information to build up the structure, thus resulting in much less efficient storage and longer parsing time compared to binary formats. In contrast, binary formats store the same information in a more compact way using some common representation of numerical data (such as the {\tt IEEE754} floating-point specification), and are preferred when large volumes of data are expected (e.g. the {\tt OUT3} file of {\tt NBODY6}, see \citealt{aarseth99}). There are, however, many binary formats for particle data, differing in how the data are arranged in the file and how it is described by the metadata (see Section~\ref{sec:formats}). Different binary formats generally produce files of similar sizes (especially when the data volume is large), and with little statistical redundancy, and therefore cannot be further reduced in size by data compression algorithms.

Big data must be written efficiently to the storage medium without interrupting or significantly slowing down the simulation process itself, thus often dedicated nodes or processors are used just to write the data to disk, while the others continue the integration (asynchronous output). Some high performance I/O libraries, such as {\tt MPI-IO}, allow multiple nodes or processors to write to the same file in parallel. Beside the writing, some ways to deal with big data in this context are utilized as needed. If the data are sorted in a certain way this could make a snapshot file smaller by not saving the particle ID. Alternatively, if the output of a tree code makes use of the space-filling (Hilbert) curve, it is easier to rapidly access spatial sub-volumes of the data \citep{springel05}. Another way is to more frequently output a subset of particles of interest, such as black holes \citep{berczik05,berczik06}. Most importantly, to reduce the amount of data that needs to be saved at least part of the analysis is carried out ``on the fly''. Regardless of the efforts of designing highly efficient data structures, some data processing on the fly may in fact be necessary. An example for such analysis is the calculation of Lagrange radii in {\tt NBODY6} \citep{aarseth99} and saving of image files of the system in {\tt PKDGRAV} \citep{jetley08}.

In this paper we propose a scalable storage scheme for $N$-body simulation data using the {\tt HDF5} high-performance data format, because of its hierarchical nature that allows us to store time-evolving hierarchical systems such as globular clusters. This paper is organized as follows: Section~\ref{sec:direct-nbody} describes the mode of operation of a direct $N$-body code; an adaptive storage scheme inspired by \cite{Farr12} for direct $N$-body simulation data and an analysis of data rate is presented in Section~\ref{sec:bts}; other possible approaches for data scaling are presented are presented in Section~\ref{sec:ScaleData}; technical concerns and benchmarks of the proposed scheme are presented in Section~\ref{benchmarks}. Finally, applications of the proposed storage scheme are presented in Section~\ref{sec:applications}.

\section{Direct $N$-body Simulations} \label{sec:direct-nbody}
In the direct $N$-body scheme, the equation of motion for a particle of index $i$ in a system containing $N$ particles takes the form \citep{aarseth03}:
\begin{equation}
	\ddot{\mathbf{r}}_{i} = -G \sum_{\substack{j=1 \\ j\neq i}}^{N} \frac{m_j(\mathbf{r}_i-\mathbf{r}_j)}{|\mathbf{r}_i-\mathbf{r}_j|^3},
\end{equation}
where $m_j$ are the masses of the other particles, $N$ is the total number of particles, $\mathbf{r}$ are the positions, and $G$ is the gravity constant. Full calculation of the mutual gravitational forces for a system of $N$ particles corresponds to $\sim N^2$ terms. The positions and velocities are updated subsequently by assuming that the evaluated force exerted on the particle is constant or can be interpolated with a polynomial during a certain time step $\Delta t$ (see more information about the integrator below). When the time step $\Delta t$ is shared among all particles, the total number of calculations for $T_\mathrm{total}$, $N$-body time step is
\begin{equation} \label{eq:naive_size}
	S = \frac{1}{2} N(N-1) \frac{T_\mathrm{total}}{\Delta t}.
\end{equation}

\begin{figure}
	\begin{center}
	\includegraphics[scale=0.45]{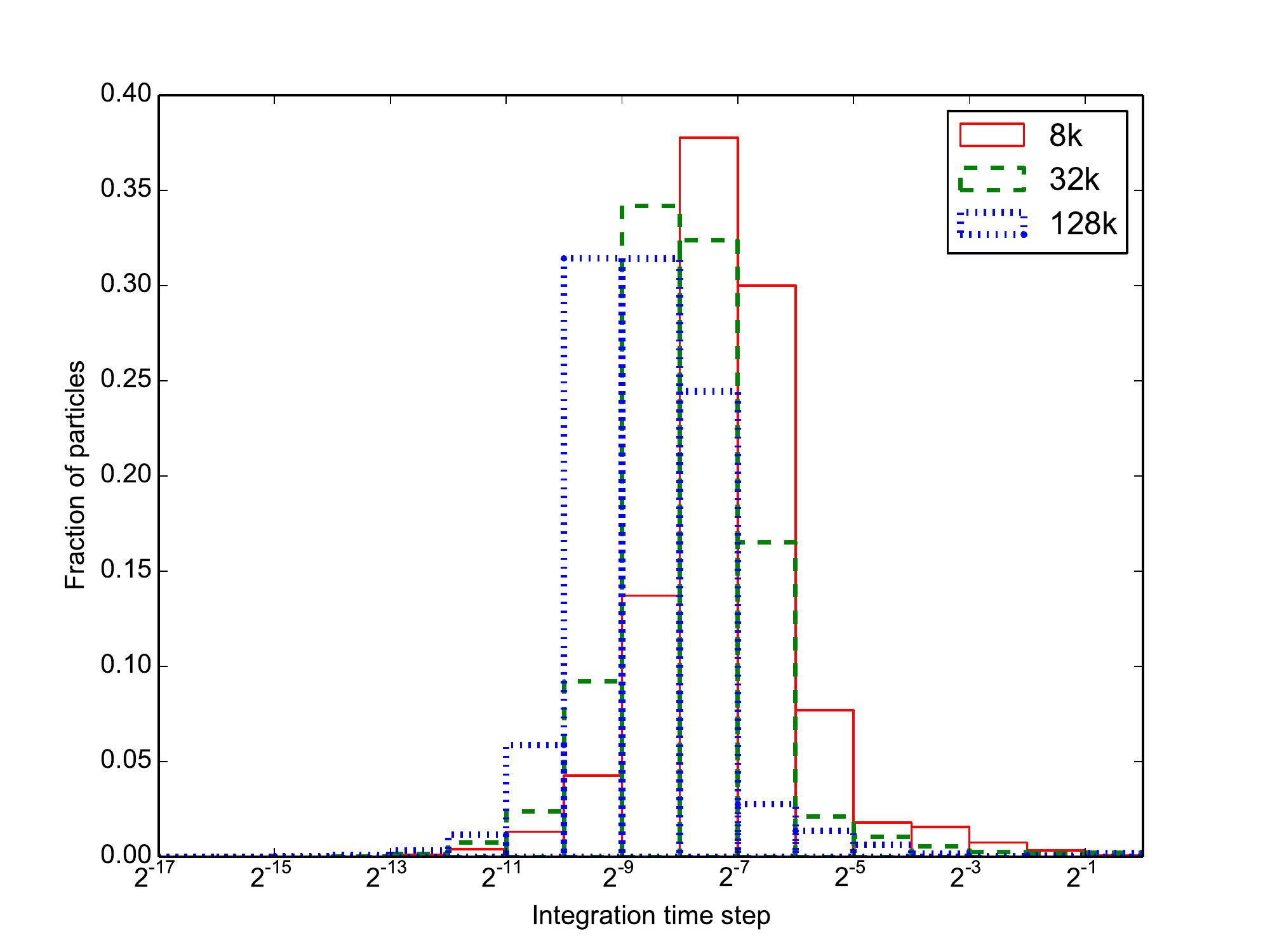}
	\caption{Time step distribution for Plummer models realizations with $N=8\mathrm{k}, 32\mathrm{k}$ and $128\mathrm{k}$ at $t=1$ H\'enon time unit, simulated with the direct $N$-body code {\tt NBODY6++}. The code employs the Block Time Step (BTS) integration scheme and a fourth order Hermite integrator. For some arbitrarily defined maximum time step $\Delta t_\mathrm{max}$, all smaller time steps are given by $\Delta t_n = \Delta t_\mathrm{max}/2^{n-1}$, where $n$ is called the ``depth of integration''. In this figure, the time steps are in H\'enon units. The peaks of the three distributions are shifted to the left as $N$ increases, illustrating that systems with higher number density have more close pairs, which lead to smaller time steps on average. }
    \label{fig:timestep_hist}
	\end{center}
\end{figure}

The choice of $\Delta t$ varies among different integration algorithms. Employing higher-order algorithms allows faster convergence, but this requires additional computational effort to calculate the high order terms \citep{acs03}. The fourth-order Hermite integrator was demonstrated to be successful in achieving an acceptable balance between accuracy and speed \citep{aarseth99b}. Nevertheless, in the dense central region of galaxies or star clusters, where close encounters may occur frequently, very small integration time steps still have to be taken in order to ensure accuracy, which will dramatically slow down the simulation. Close encounters will eventually cause tight binary systems to form: such systems need permanent treatment with extremely small time steps. In this almost inevitable scenario, should all integration points be saved, the sizes of the resulting data files would be overwhelming. For instance, in a globular cluster with $N=10^5$, if the system is to be evolved for 1000 H\'enon time units\footnote{The $N$-body unit system is referred to here as the H\'enon unit system in honor of Michel H\'enon.} with $\Delta t=10^{-4}$ (which is relatively large; see histogram in Fig.~\ref{fig:timestep_hist}), more than $10^{18}$ bytes (1 exabyte) of data would be generated in total. This would pose great challenges even for modern storage arrays and for data analysis.

As direct computation of the $\mathcal{O}(N^2)$ algorithm is expensive, optimization schemes such as \emph{individual time step} (ITS), \emph{Ahmad-Cohen neighbor scheme} (ACS; \citealt{ahmad73}) were developed to dramatically reduce the computational costs \citep{aarseth03}. Almost all modern $N$-body integrators now employ the ITS scheme. The basic idea is that since gravity follows an inverse-square law, particles from regions of different density experience different magnitudes of force. The density profiles of globular clusters can be roughly approximated by a power law, such as the Plummer model \citep{plummer} or the King model \citep{king67}. Stars in the outskirts of star clusters usually move relatively unperturbed for timescales comparable to hundreds of times the corresponding timescales of the central particles, and hence long time steps can be used for their integration. Stars in the central regions, however, frequently experience violent interactions (close encounters) with their neighbors and therefore require much smaller integration time steps. Integration of a particle with index $i$ is therefore carried out using a time step $\Delta t_i$, which is often taken to be \citep[e.g.][]{aarseth03}
\begin{equation} \label{eq:timestep_criterion}
	\Delta t_{i} = \sqrt{\eta \frac{|\mathbf{a}_{i}|
            |\mathbf{a}^{(2)}_{i}| +
            |\dot{\mathbf{a}}_{i}|^2}{|\dot{\mathbf{a}}_{i}|
            |\mathbf{a}^{(3)}_{i}| +
            |\mathbf{a}^{(2)}_{i}|^{2}}},
\end{equation}
where $\mathbf{a}_i$ is the acceleration of particle $i$ (the total force acting on it divided by its mass), $\dot{\mathbf{a}}_i$, $\mathbf{a}_i^{(2)}$ and $\mathbf{a}_i^{(3)}$ are the first, second and third derivatives of the acceleration. The parameter $\eta$ controls the accuracy of the integration and a commonly used value is $\eta=0.02$ \citep{aarseth03}. Depending on different density profiles, the ITS scheme reduces the computational complexity from $\mathcal{O}(N^2)$ to $\mathcal{O}(N^{4/3})$, and a larger gain can be achieved with centrally concentrated systems \citep{makino88}. Here, a particle is considered as \emph{active} if its state is changed significantly in time scales comparable to the integrator time step. Fig.~\ref{fig:timestep_hist} shows a time step distribution of time steps for systems with $N=8\mathrm{k},32\mathrm{k}$ and $128\mathrm{k}$ (in this paper $\mathrm{k} = 2^{10} = 1024$).

\begin{figure}
	\begin{center}
	\includegraphics[scale=0.45]{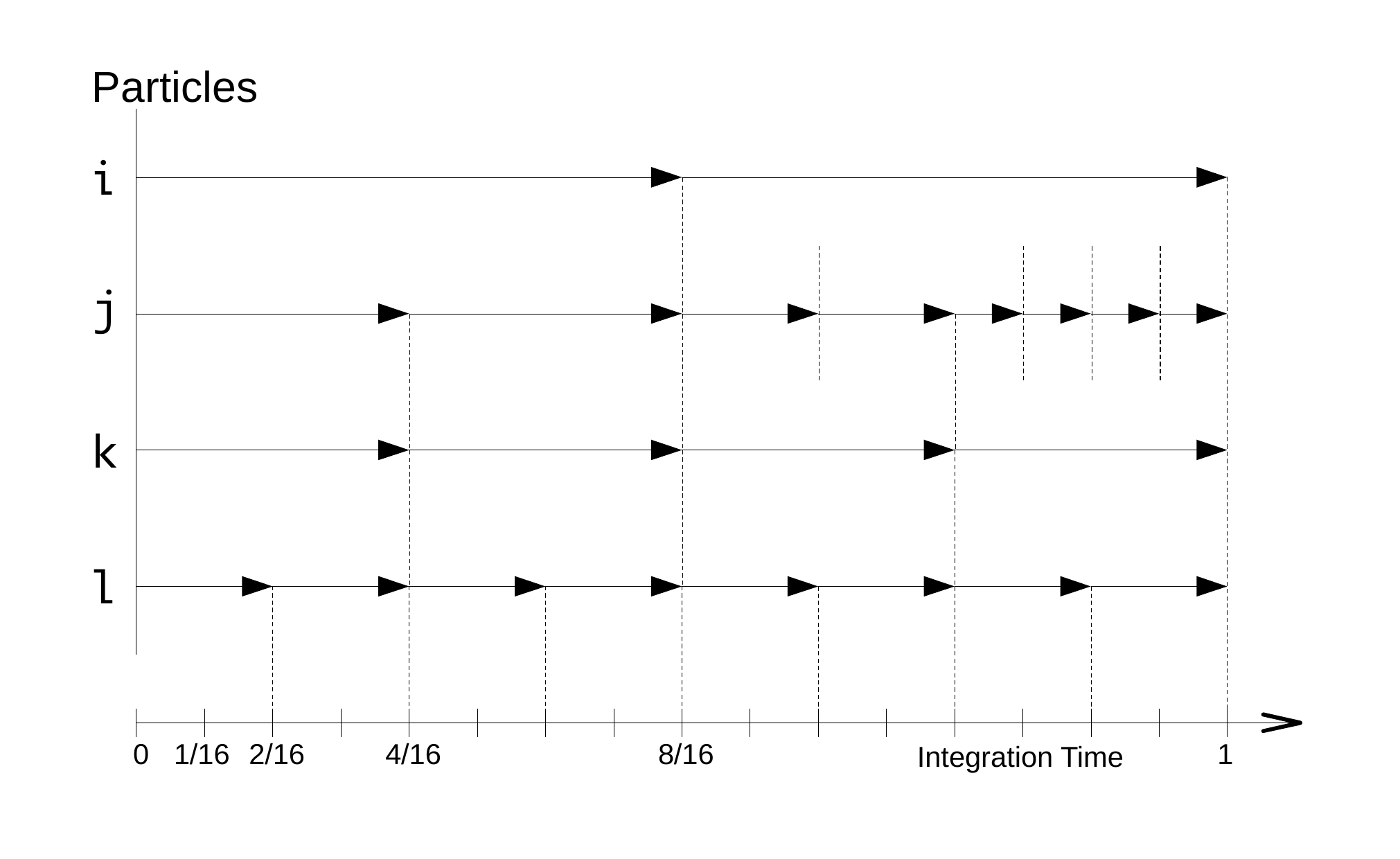}
	\caption[dd]{Schematic illustration of a 4-particle system integrated with a block time step (BTS) scheme. Particles $(i, j, k, l)$ are assigned individual time steps according to the forces exerted on them. It is assumed here that the total integration time is 1 (in arbitrary unit) and the minimum integration time is $\nicefrac{1}{16}$. At $t=\nicefrac{1}{16}$, no particle is scheduled to be integrated, as none have time step smaller than $\Delta t=\nicefrac{2}{16}$. As the system proceeds to $t=\nicefrac{2}{16}$, particle $l$ is the only particle with a time step short enough to schedule an integration. At $t=\nicefrac{4}{16}$, particles $(j, k, l)$ are scheduled for integration while particle $i$ is still outside the list. The full system is integrated at $t=\nicefrac{8}{16}$. Since after that particle $j$ becomes increasingly active, it is integrated every time step starting from $t=\nicefrac{12}{16}$. The BTS scheme assigns time step in a hierarchical fashion based on equation~(\ref{eq:block_time_step}), and therefore guarantees that the commensurability of the individual time step of all particles. }
	\label{fig:block_time_step}
	\end{center}
\end{figure}

According to equation~(\ref{eq:timestep_criterion}), the time step $\Delta t_i$ of particle $i$ can get an arbitrary value. In practice, however, in order to divide particles into groups according to their time steps, the \emph{block time step} (BTS) scheme is often employed, permitting particles in the same time step group to be advanced at the same time \citep{hayli67,hayli74,mcmillan86}. Fig.~\ref{fig:block_time_step} illustrates how particles are advanced in the BTS scheme. For instance, in the hierarchical scheme used by {\tt NBODY6} and its parallel version {\tt NBODY6++}\footnote{This paper makes no distinction between {\tt NBODY6} and {\tt NBODY6++}.} \citep{spurzem99,spurzem08} as well as many other Aarseth-type codes, the time steps are defined as
\begin{equation} \label{eq:block_time_step}
	\Delta t_n = \Delta t_\mathrm{max} / 2^{n-1},
\end{equation}
where $n$ is the level of integration and $\Delta t_\mathrm{max}$ is a predefined maximum time step (which in practice is often taken as one H\'enon time unit).

The integration itself is often done using a predictor-corrector scheme. As an example, the Hermite Scheme employed by {\tt NBODY6++} first predicts the positions $\mathbf{x}_{i,p}(t)$ and the velocities $\mathbf{v}_{i,p}(t)$ at some time $t$:
\begin{align}
  \mathbf{x}_{i,p}(t) &= \mathbf{x}_{i,0} + (t - t_0)\mathbf{v}_{i,0}
                       + \frac{\left(t - t_0 \right)^2}{2}\mathbf{a}_{i,0}
                       + \frac{\left(t - t_0 \right)^3}{6}\mathbf{\dot{a}}_{i,0}\label{Hermite_x}\\
  \mathbf{v}_{i,p}(t) &= \mathbf{v}_{i,0} + (t - t_0)\mathbf{a}_{i,0}
                       + \frac{(t-t_0)^2}{2}\mathbf{\dot{a}}_{i,0}\label{Hermite_v}
\end{align}
where $t_0$ is the starting time and the subscript 0 of the vector quantities denotes the known value of the quantity at $t_0$. The acceleration $\mathbf{a}_{i}$ and its first derivative $\dot{\mathbf{a}}_{i}$ are evaluated at time $t$ at the predicted position (i.e. by direct summation), and the two higher order derivatives of the acceleration can be evaluated at time $t_0$:
\begin{align}
  \mathbf{a}^{(2)}_{i,0} &= -6\frac{\mathbf{a}_{i,0}-\mathbf{a}_{i}}{(t-t_0)^2}
                          -  2\frac{\mathbf{\dot{a}}_{i,0}
                          + \mathbf{\dot{a}}_{i}}{t-t_0}\label{a_2}\\
  \mathbf{a}^{(3)}_{i,0} &= 12\frac{\mathbf{a}_{i,0}-\mathbf{a}_{i}}{(t-t_0)^3}
                          + 6\frac{\mathbf{\dot{a}}_{i,0}+\mathbf{\dot{a}}_{i}}{(t-t_0)^2}\label{a_3}
\end{align}
The second and third derivatives can be used to correct the predicted values to fourth order:
\begin{align}
  \Delta \mathbf{x}_{i} &= \frac{1}{24}\mathbf{a}^{(2)}_{i,0}(t - t_0)^4
                         + \frac{1}{120}\mathbf{a}^{(3)}_{i,0}(t - t_0)^5\label{Hermite_pred_x}\\
  \Delta \mathbf{v}_{i} &= \frac{1}{6}\mathbf{a}^{(2)}_{i,0}(t - t_0)^3
                         + \frac{1}{24}\mathbf{a}^{(3)}_{i,0}(t - t_0)^4\label{Hermite_pred_v}
\end{align}
Finally, the corrected position $\mathbf{x}_{i}(t)$ and velocity $\mathbf{v}_{i}(t)$ at the time $t$ can be expressed as
\begin{align}
  \mathbf{x}_{i}(t) &= \mathbf{x}_{i,p}(t) + \Delta \mathbf{x}_{i}\label{Hermite_x_corrected}\\
  \mathbf{v}_{i}(t) &= \mathbf{v}_{i,p}(t) + \Delta \mathbf{v}_{i}\label{Hermite_v_corrected}
\end{align}
Therefore, extra terms $\mathbf{a}_{i,0}$ and $\mathbf{\dot{a}}_{i,0}$ need to be stored, and interpolation also introduces extra computational overhead. For certain applications, such as visualization, it may not be critical to compute the corrector terms, so part of the computational and storage overhead can be further reduced.

\section{BTS Storage Scheme}\label{sec:bts}

\subsection{Description}

Originally inspired by the ITS scheme, \cite{Farr12} have shown that the data can actually be significantly compressed by recording only active particles. Below, we estimate the data rate of this approach by first considering the ``traditional'' snapshot scheme. During one H\'enon time unit, the number of data records produced by the scheme is
\begin{equation} \label{eq:size_snapshot}
	\textsc{Size(Snapshots)} = 2^{R_t} N,
\end{equation}
where $R_t$ is the temporal resolution factor, such that in one H\'enon time unit, the output operation is triggered for $2^{R_t}$ times. In the BTS scheme,
\begin{align}
	\textsc{Size(BTS)} &= \sum\limits_{n=0}^{{R_t}-1} 2^n N_n + 2^{R_t} \sum\limits_{n={R_t}}^{\infty} N_n\nonumber\\
	&= \sum\limits_{n=0}^{{R_t}-1} 2^n N_n + 2^{R_t}\left[N-\sum\limits_{n=0}^{{R_t}-1} N_n \right],\label{eq:size_bts}
\end{align}
where $N$ is the total number of particles and $N_n$ is the number of particles with time step $\Delta t = 1/2^n$. For a given ${R_t}$, particles with integration time step $\Delta t_i \geq 1/2^{R_t}$ are fully resolved, in the sense that output is commensurate with integration. That is, whenever these particles are integrated (or in the terminology of the Hermite scheme, corrected) their data are written to the file; moreover, this happens \emph{only} when integration is performed. The rest of the particles, which have $\Delta t_i < 1/2^{R_t}$, are not fully resolved (for particles with time step $1/2^n$, output occurs only every $2^{n-{R_t}}$ integrations).

In the snapshot scheme, since data from all particles are written at the same time, particles with integration frequencies lower than the output frequency have to be extrapolated (or in the terminology of the Hermite scheme, predicted); this is redundant, since analysis software can do this prediction, which is computationally very cheap. The BTS scheme compresses the data of the fully resolved particles by eliminating redundant information, which is \emph{lossless}. The BTS scheme compresses the data of particles with integration frequency higher than the output frequency by skipping a certain number of integration points, which is \emph{lossy}.

\begin{figure}
	\begin{center}
	\includegraphics[scale=0.45]{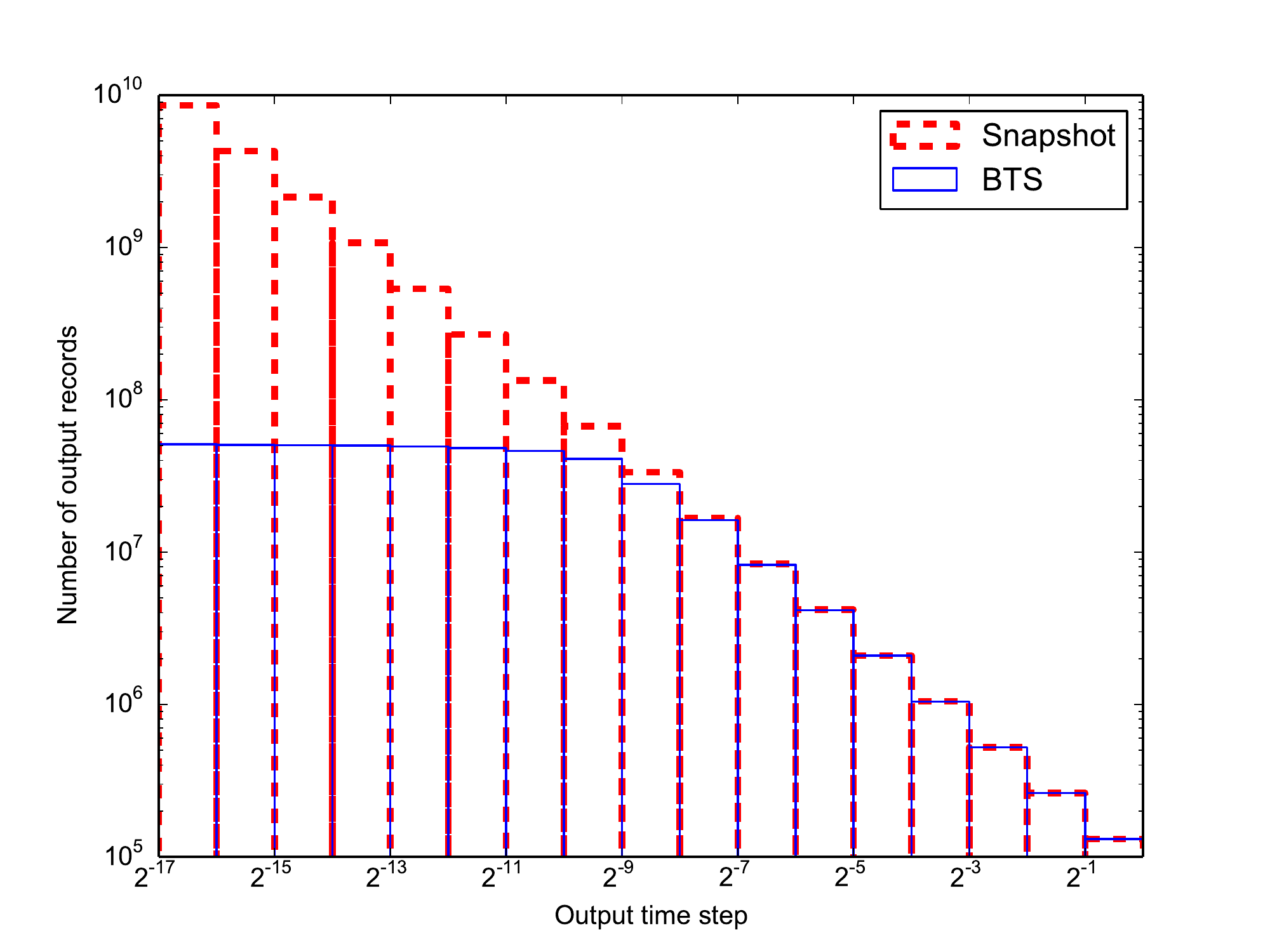}
	\caption{
	According to the time step distribution of the $N=128\mathrm{k}$ simulation in Fig.~\ref{fig:timestep_hist} at $t=1$, these histograms show the number of output records as a function of output time step (in H\'enon units), using the BTS scheme (solid blue histogram) and the snapshot scheme (dashed red histogram). }
	\label{fig:snapshot_vs_bts}
	\end{center}
\end{figure}

Fig.~\ref{fig:snapshot_vs_bts} shows that the BTS file size initially grows exponentially as a function of ${R_t}$ (like the snapshot scheme) but turns over at output frequency close to the peak of the time step distribution (in Fig.~\ref{fig:timestep_hist}) and saturates (so that the file size remains finite even when the output frequency grows to infinity, on the left of the figure). This saturation is due to the small number of particles with very small time steps, as seen in Fig.~\ref{fig:timestep_hist}. Note that for very low output frequencies, the snapshot and BTS files have similar sizes, converging at ${R_t}=0$ (which in this case represents the maximally allowed time step).

The snapshot scheme is prohibitively expensive if one intends to resolve the most rapidly varying particles, but this is feasible due to the convergence property of the BTS scheme. On the other hand, for low output frequencies, the methods are equivalent in terms of number of records. The snapshot scheme might even be preferable in this case since the extra particle attributes $\mathbf{a}_{0}$ and $\mathbf{\dot{a}}_{0}$ need not be stored.

\subsection{Example}\label{sec:temporal}

Consider as an example the 4-particle system illustrated in Fig.~\ref{fig:block_time_step}, where the system is integrated for one H\'enon time unit with minimum integration time step $\Delta t = \nicefrac{1}{16}$; a temporal resolution factor of $R_t=3$ is adopted, which means that $2^{R_t}=8$ output operations are scheduled within this one H\'enon time unit. The first output (not including $t=0$) is triggered at $t=\nicefrac{1}{8}$, when particle $l$ is the only particle in the output list; at $t=\nicefrac{2}{8}$ however, particles $(j,k,l)$ are active and eligible for output. At $t=\nicefrac{11}{16}$, although particle $j$ is integrated, no output occurs as it is not a product of the output time step $2^{-R_t}$ and an integer, therefore the information is lost here. It is instead included in the output list at $t=\nicefrac{12}{16}=\nicefrac{6}{8}$ alone with particle $l$, and only the latest data at $t=\nicefrac{12}{16}$ will be written. At $t=\nicefrac{8}{16}$ and $t=\nicefrac{1}{8}$ the system receives a full output.

\begin{figure}
	\begin{center}
	\includegraphics[scale=0.45]{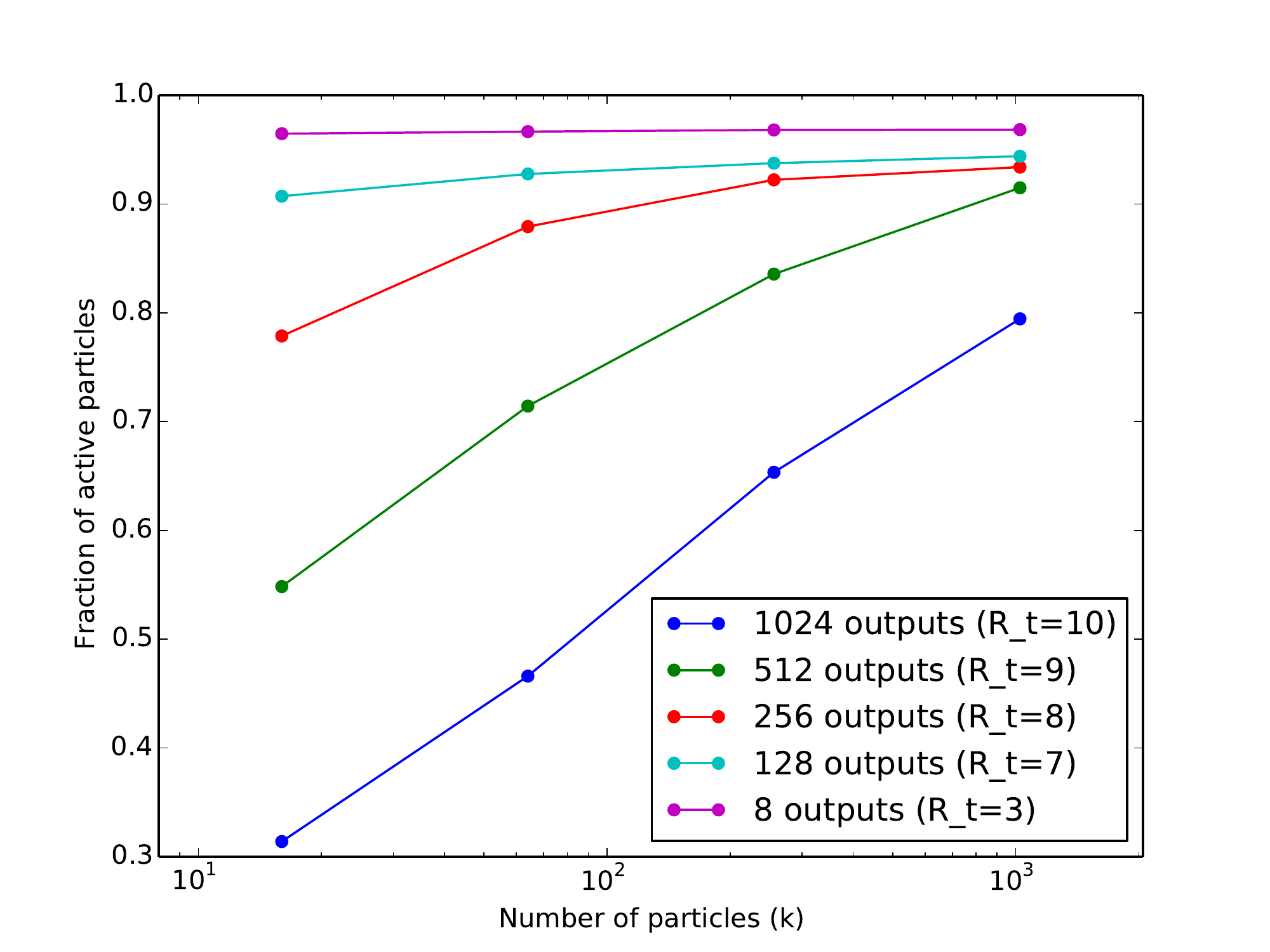}
	\caption{Fraction of active particles averaged over one H\'enon time unit. Plummer systems with $N=16\mathrm{k},64\mathrm{k},256\mathrm{k}$ and $1024\mathrm{k}$ are evolved, and with each $N$ five different temporal resolutions $R_t$ corresponding to five different output frequencies are marked with different colors of lines. As the output frequency goes higher, the fraction of active particles declines, and therefore the BTS storage scheme gains significant reduction of data rates. For the same output frequency, systems with larger $N$ have higher fractions of active particles. The two methods converge at $R_t=0$, which yields standard snapshots.}
	\label{fig:frac_active_particles}
	\end{center}
\end{figure}

At any time the number of particles in the output list will not exceed the total number of particles. Consider again the above example while $R_t=2$, more particles will be collected at the output points but the output interval is longer. Recall that for a full snapshot output, according to equation~(\ref{eq:size_snapshot}) the size of output is proportional to $N$, the output size of this scheme is a \emph{linear} function of $N$. Statistics of the active particle fractions for simulations carried out with {\tt NBODY6++} are presented in Fig.~\ref{fig:frac_active_particles}. The linear property of the BTS scheme makes the scheme suitable for very large systems, as long as the detailed evolution of the highly active particles is not important. For example, the resulting datasets can be used to generate visualization data for the overall evolution of star clusters. The datasets will have sufficient resolution to describe slow particles in the outskirts of the cluster in detail, allowing the viewers to observe the evaporation process. Since the output frequencies of highly active particles have been truncated to $2^{R_t}$, the datasets will only have enough resolution for these particles if the output resolution $R_t$ is set to sufficiently high (so that the output time step is comparable with the actual integration of those particles). Even in such case, the output size will still converge to a manageable scale as seen from Fig.~\ref{fig:snapshot_vs_bts}.

\subsection{Interpolation}\label{sec:splines}

Analysis or visualization software needs to interpolate the positions and sometimes velocities and higher derivatives of the particles between the output points. This is best done using septic splines, which are seventh degree piecewise polynomials. This method ensures that the interpolated curves exactly touch at the endpoints (or the spline's knots) and that  the stored information about the previous and next known states is used. Lower order interpolation can be used based on the predictor-corrector scheme (Section~\ref{sec:direct-nbody}) or by applying a lower order splines that would discard some stored information (i.e. the known values of $\dot{\mathbf{a}}$); higher order interpolation can be achieved if one uses more than the two nearest points. Let us define the running variable:
\begin{equation}
  \tau \equiv \frac{t - t_0}{t_1 - t_0} = \frac{t - t_0}{\Delta t}
\end{equation}
such that $0 < \tau < 1$, where $t$ is an arbitrary time in which we are interested in the particle's properties; $t_0$ is the last integration point before time $t$, and $t_1$ is the next one; $\Delta t \equiv t-t_0$ is the output time step. The interpolated position of the particle is thus:
\begin{equation}
\mathbf{x}(\tau) = \mathbf{p}_{0}+\mathbf{p}_{1}\tau+\mathbf{p}_{2}\tau^{2}+\mathbf{p}_{3}\tau^{3}+\mathbf{p}_{4}\tau^{4}+\mathbf{p}_{5}\tau^{5}+\mathbf{p}_{6}\tau^{6}+\mathbf{p}_{7}\tau^{7}
\end{equation}
where $\mathbf{p}_{0}\ldots\mathbf{p}_{7}$ are the spline coefficients given in equations (\ref{eq:spline0}) to (\ref{eq:spline7}). The expressions for the velocity and higher derivatives can be easily determined from the above expression by derivation. Let the subscripts 0 and 1 represent the values of the quantities at times $t_0$ and $t_1$ respectively, so we can write:
\begin{align}
\mathbf{p}_{0} &= \mathbf{x}_{0}\label{eq:spline0}\\
\mathbf{p}_{1} &= \mathbf{v}_{0}\Delta t\\
\mathbf{p}_{2} &= \textstyle{\frac{1}{2}}\mathbf{a}_{0}\Delta t^{2}\\
\mathbf{p}_{3} &= \textstyle{\frac{1}{6}}\dot{\mathbf{a}}_{0}\Delta t^{3}\\
\mathbf{p}_{4} &= -\textstyle{\frac{1}{6}}\left(4\dot{\mathbf{a}}_{0}+\dot{\mathbf{a}}_{1}\right)\Delta t^{3}-\textstyle{\frac{5}{2}}\left(2\mathbf{a}_{0}-\mathbf{a}_{1}\right)\Delta t^{2}\nonumber\\
 & -5\left(4\mathbf{v}_{0}+3\mathbf{v}_{1}\right)\Delta t-35(\mathbf{x}_{0}-\mathbf{x}_{1})\\
\mathbf{p}_{5} &= \textstyle{\frac{1}{2}}\left(2\dot{\mathbf{a}}_{0}+\dot{\mathbf{a}}_{1}\right)\Delta t^{3}+\left(10\mathbf{a}_{0}-7\mathbf{a}_{1}\right)\Delta t^{2}\nonumber\\
 & +3\left(15\mathbf{v}_{0}+13\mathbf{v}_{1}\right)\Delta t+84(\mathbf{x}_{0}-\mathbf{x}_{1})\\
\mathbf{p}_{6} &= -\textstyle{\frac{1}{6}}\left(4\dot{\mathbf{a}}_{0}+3\dot{\mathbf{a}}_{1}\right)\Delta t^{3}-\textstyle{\frac{1}{2}}\left(15\mathbf{a}_{0}-13\mathbf{a}_{1}\right)\Delta t^{2}\nonumber\\
 & -2\left(18\mathbf{v}_{0}+17\mathbf{v}_{1}\right)\Delta t-70(\mathbf{x}_{0}-\mathbf{x}_{1})\\
\mathbf{p}_{7} &= \textstyle{\frac{1}{6}}\left(\dot{\mathbf{a}}_{0}+\dot{\mathbf{a}}_{1}\right)\Delta t^{3}+2\left(\mathbf{a}_{0}-\mathbf{a}_{1}\right)\Delta t^{2}\nonumber\\
 & +10\left(\mathbf{v}_{0}+\mathbf{v}_{1}\right)\Delta t+20(\mathbf{x}_{0}-\mathbf{x}_{1})\label{eq:spline7}
\end{align}

Note that the discussion above is \emph{per particle}, and that each equation represents three vector components, which for the purpose of the interpolation are completely independent.

\section{Modified BTS storage schemes}\label{sec:ScaleData}

As shown in Fig.~\ref{fig:snapshot_vs_bts}, the BTS file size converges with increasing output frequency, and therefore this scheme becomes mandatory for the storage of simulation data when very high temporal resolution is required. Nevertheless, BTS scheme integration data of a very large simulation can still be too large even for moderately large systems. While equation~(\ref{eq:size_bts}) shows that that the data are scalable by specifying an output frequency $R_t$, this scaling technique may not provide sufficient resolution for highly active particles. An alternative scaling technique is presented in Section~\ref{sec:spatial}, allowing the user to define an individual resolution for each particle. While dynamically active particles are assumed to be interesting particles, Section~\ref{sec:particle_interest} explores some possible scenarios where this may not necessarily be the case. Finally, Section~\ref{sec:attrib_driven} discusses an even more generic scenario in which the output may be driven by physical processes other than the dynamical evolution.

\subsection{Scaling with Spatial Resolution: Triggered Output for Significantly Updated Particles} \label{sec:spatial}
In this scheme, the output is triggered \emph{per particle}: whenever an individual particle has been integrated $R_s$ times, its data is eligible for output. Since $R_s$ defines the portion of integration for output, it defines the resolution in a spatial manner: larger values of $R_s$ correspond to lower spatial resolution, and $R_s=1$ corresponds to a full output of all BTS integration data. For $R_s > 1$, the scheme skips $R_s - 1$ integrations right after the current output until the next one, reducing the data rate by a factor comparable to $R_s$ (the time steps of all particles are changing as they move in phase space and therefore it is unlikely that the reduction factor is exactly $R_s$). The total output rate is proportional to the total number of individual time steps, and the resulting output size grows as $N^{4/3}$ \citep{makino88}.

For instance, consider again the 4-particle system illustrated in Fig.~\ref{fig:block_time_step}. Assuming that $R_s = 2$, the scheme skips one output right after the current output, so particle $i$ will be only eligible for output at $t=1$; particle $j$ is eligible for output at $t=\nicefrac{8}{16}, \nicefrac{12}{16}, \nicefrac{14}{16}, 1$, and so on. With a careful choice of $R_s$, the scheme yields datasets with sufficient resolution for highly active particles such as hard binaries and close encounters, but also reduces the data rate of slow particles. Full output corresponds to $\sim1000$ records per particle orbit (on average); for rendering purposes it is still sufficient to reduce this by one order of magnitude, and hence reducing the storage consumption by one order of magnitude. A comparison of the file sizes for different values of $R_s$ is presented in Table~\ref{tab:NVIS}. It is sensible to apply this scheme for detailed follow-ups of energetic subsystems.

\begin{table}
\begin{center}
\caption{\label{tab:NVIS}Spatial output resolution $R_s$ as a function of number of records.}
\begin{tabular}{cccc}
\hline
\hline
 $R_s$ &  \# of records (w/ BTS) &  \# of records (w/o BTS) &  Efficiency ratio \\
\hline
50 & 112128 & 536870912 & 4788.0 \\
40 & 223934 & 536870912 & 2343.0 \\
30 & 272236 & 536870912 & 1972.1 \\
20 & 404532 & 536870912 & 1327.1 \\
10 & 766584 & 536870912 & 700.3 \\
1 & 6953525 & 536870912 & 77.2 \\
\hline
\end{tabular}
\end{center}
\textbf{Notes.} The simulation was carried out with {\tt NBODY6++} for one H\'enon time unit (roughly 1 Myr) and $N=16384$ particles, where BTS is employed. $R_s = 1$ corresponds to the full output of BTS data. The smallest time step of an $N=16384$ Plummer system is of the order of $\Delta t \sim 2^{-15}$, according to the time step distribution given by Fig.~\ref{fig:timestep_hist}. Hence, should there be no BTS scheme, the total number of record is proportional to $N^2\Delta t \sim 5.3 \times 10^8$ according to equation~(\ref{eq:naive_size}) (shown in column 2). A full output of BTS data already yielded an efficiency ratio (column 2 divided by column 1) of 77.2, and together with the $R_s$ parameter the reduction can be promising.
\end{table}

\subsection{Dedicated Output for the Particles of Interest} \label{sec:particle_interest}
Binary and triple black holes in galactic nuclei or star cluster centers, hypervelocity stars and the host stars of planetary systems are particularly interesting objects to investigate in simulations. The output module of the integrator should therefore accommodate this need by providing high-resolution output for the particles of interest (POIs) while suppressing the output of uninteresting particles to achieve maximum storage efficiency. POIs can be dynamically active. For example, a binary black hole system in a galactic nucleus can be so dynamically active that it will take a significant fraction of the wall-clock time to resolve even with advanced regularization technique (e.g. KS regularization). However, the bouns of these computations is that they keep the data of those active particles up-to-date all the time, allowing the output module to simply dump the data without interpolation. On the other hand, it may be interesting to follow the evolution of hypervelocity binary stars in the outskirts of the cluster (e.g. \citealt{lu07}). The forces exerted on those objects change rather slowly, despite their high velocities. Consequently, the integrator will not integrate those objects frequently; reliable dynamical data can then be achieved with interpolation, for example, with the spline method presented in Section~\ref{sec:splines}.

\subsection{Events/Attributes Driven Output} \label{sec:attrib_driven}
The output scenarios previously discussed are all driven by the dynamical evolution of the simulated system. Output will be triggered when the coordinates of particles change significantly. Sometimes, however, it is necessary to have the output triggered by certain events and/or attributes. For example, in starburst galaxies, the star formation process is usually the most interesting process to investigate. Critical events of stellar evolution may not necessarily correspond to critical events of the dynamical evolution. Therefore, output strategy should instead be driven by the stellar evolution process, allowing the follow-up data analysis to trace the evolution of such astrophysical processes.

Gravitational dynamics codes generally provide dynamical information of the particles such as positions, velocities and accelerations. Some codes support simulation of multiple astrophysical processes simultaneously. For example, {\tt NBODY6++} is able to take the feedback of stellar evolution into account while handling the dynamical processes of particles. For direct $N$-body code with stellar evolution, possible assignment of astrophysical quantities for individual particles could be tabulated as in Table~\ref{table:DataStructure}; binary systems are very common in such simulations, which requires auxiliary data structure to describe their properties as a whole. A possible binary data structure is presented in Table~\ref{table:BinaryDataStructure}.

\begin{table}
\begin{center}
\caption{Astrophysical quantities of individual particles in a direct $N$-body simulation with stellar evolution}
\begin{tabular}{ccc}
\hline
\hline
Quantity & Meaning & Category  \\
\hline
$i$ & Unique identifier of the particle & Miscellaneous	 \\
{\tt name} & User friendly label (e.g. for visualization) & Miscellaneous \\
$t$ & Current time & Miscellaneous \\
$\delta t$ & Next time step & Miscellaneous \\
$m$ & Mass & St. dyn. \& evo. \\
$x$ & Position vector & Stellar dynamics \\
$\dot{x}$ & Velocity vector & Stellar dynamics \\
$a$ & Acceleration vector & Stellar dynamics \\
$\dot{a}$ & Jerk vector (first derivative of $a$) & Stellar dynamics \\
$\rho$ & Neighbor density & Stellar dynamics \\
$\phi$ & Local potential & Stellar dynamics \\
$t_\textsc{ev}$ & Stellar evolution age & Stellar evolution \\
$k_{\textsc{star}}$ & Type indicator of star & Stellar evolution \\
$L$ & Luminosity & Stellar evolution \\
$R$ & Radius & Stellar evolution \\
$T_\textsc{eff}$ & Effective temperature & Stellar evolution \\
$Z$ & Metallicity & Stellar evolution \\
$\delta m$ & Mass change during $t_\textsc{ev}$ & Stellar evolution \\
$m_\textsc{core}$ & Core mass & Stellar evolution \\
$r_\textsc{core}$ & Core radius & Stellar evolution \\
\hline
\end{tabular}
\end{center}
\label{table:DataStructure}
\end{table}

\begin{table}
\begin{center}
\caption{Astrophysical quantities of binary systems in a direct $N$-body simulation.}
\begin{tabular}{ccc}
\hline
\hline
Quantity & Meaning  \\
\hline
$i_1, i_2$ & Unique identifiers of the two particles \\
$P$ & Orbital period  \\
$A$ & Semi-major axis \\
$e$ & Eccentricity of the binary orbit \\
$I$ & Orbital inclination \\
$I_1, I_2$ & Inclinations of the spins\\
$x_c$ & Position vector of the center of mass \\
$\dot{x_c}$ & Velocity vector of the center of mass \\
\hline
\end{tabular}
\end{center}
\textbf{Notes.} Individual properties of each component can be retrieved by referring to Table~\ref{table:DataStructure} with $i_1$ and $i_2$.
\label{table:BinaryDataStructure}
\end{table}

\section{Technical Concerns and Benchmarks}
\label{benchmarks}
Even on modern supercomputers, a physically realistic simulation may take months to run. It is therefore critical to store the output such that it is accessible for further analysis. This requires that data be stored in a well behaved, high performance data structure. Generally, a simulation data file should meet the following requirements:
\begin{itemize}
	\item Accuracy: correctly recording the relevant quantities;
	\item Time efficiency: data are written faster than they are generated, and the simulation is not slowed down significantly due to data output;
	\item Space efficiency: redundancy minimized;
	\item Interchangeability: machine/OS independent;
	\item Scalability: scalable to simulations big and small, simple and complicated; and
	\item Robustness: data loss minimized when the file is corrupted.
\end{itemize}

Datasets of $N$-body simulations are designed to describe a time-evolving system. Since inactive particles are not recorded, the interpolation of their data requires knowledge of their previously active state, making the dataset itself time-dependent. Hence, guarantee of data consistency would be another requirement.

\subsection{Choosing a File Format} \label{sec:formats}
File formats are roughly divided into two categories: \textsc{ascii} files and binary files. \textsc{ascii} files are generally easier to interpret and are human-readable. For example, tabular data are often stored as {\tt CSV} (comma-separated values) files. Since the {\tt CSV} data format simply uses one delimiter character (e.g. a comma) to separate fields, and uses a line break to indicate the termination of a record, it is widely supported and can be easily imported to an analysis program. More complicated \textsc{ascii} or text formats, such as {\tt XML}\footnote{\tt http://www.w3.org/XML/} and {\tt YAML}\footnote{\tt http://www.yaml.org/} also have been developed and standardized, making text files capable of describing hierarchical data structures. Text files avoid some of the problems encountered with binary files, such as endianness, padding bytes, and differences in the number of bytes in a machine word. However, they are not native to computer systems. Indeed, representation of numerical values in a text file is just literal, as these values are merely \textsc{ascii} sequences, and have to be converted to their intrinsic values before any computation can be performed. Standardized text formats, such as {\tt XML}, structure the data with tags, which contributes to its low entropy.

In contrast, high I/O throughput are usually achieved with binary files, since they are byte sequences native to the machines. High level binary file libraries have been developed to resolve the problems of endianness, padding bytes, file headers, metadata storage, block data storage, etc. The {\tt HDF5}\footnote{\tt http://www.hdfgroup.org/} (Hierarchical Data Format, version 5) for example, is an implementation of a binary file standard dedicated to handling large volumes of numerical data. It offers rich features such as compression filters, checksum filters, chunking, partial I/O, parallel I/O and caching. It allows the data to be structured in a hierarchical fashion and being accessed using {\tt POSIX}-like path syntax. While the {\tt HDF5} format is designed for general purpose numerical data storage, some higher level application programming interfaces have been developed to fit into special applications. For example, {\tt H5Part} \citep{Adelmann08} is a portable high performance parallel data interface for {\tt HDF5}, which is dedicated for the storage of particle-based simulation data. Other widely used binary file formats in astrophysics include {\tt CDF}\footnote{\tt http://cdf.gsfc.nasa.gov/} (Common Data Format), {\tt NetCDF}\footnote{\tt http://www.unidata.ucar.edu/software/netcdf/} (Network Common Data Format) and {\tt FITS}\footnote{\tt http://fits.gsfc.nasa.gov/} (Flexible Image Transport System). All these formats are self-describing and machine-independent, optimized for scientific data. The {\tt FITS} data is mainly designed for image data as its name indicates, and the image metadata is stored in human readable \textsc{ascii} head, allowing an interested user to easily examine the header information with a simple text editor. {\tt CDF} and {\tt NetCDF} are more general data formats. Originally they share the same conceptual model based on a multidimensional (array) model, but the latter has since diverged and is not compatible with the former. Some data formats are developed and optimized for more specific applications. For instance, the {\tt SDF} format \citep{warren13} is used in the oct-tree based ``Dark Sky'' cosmological Simulations \citep{skillman14}.

\subsection{Benchmarks}
We adopt {\tt HDF5} as the native output format for the direct $N$-body code {\tt NBODY6} and {\tt NBODY6++}, due to the rich features it offers and especially its interchangeability within the astronomical community. For example, the {\tt GADGET2}\footnote{\tt http://www.mpa-garching.mpg.de/gadget/} code (which was used, among others, in the Millennium Simulation mentioned above) has an options to output its snapshot data in {\tt HDF5} format; the Low Frequency Array (LOFAR) chose to use {\tt HDF5} to manage astronomical radio data \citep{anderson11}. The internal file layout is structured with the {\tt H5Part} scheme \citep{Adelmann08}. For the purpose of benchmarks, we also store the data as plain text CSV files, in which the particle data is described by multiple columns separated by commas, and each line describes the full data for a particle. Each floating point number takes 8 bytes. Since the {\tt CSV} format is not hierarchical, the time variable for particles in the same time group is repeated many times, as shown below:
{\tt
\begin{lstlisting}
t1, 1, x1, y1, z1, vx1, vy1, vz1, ...
t1, 2, x2, y2, z2, vx2, vy2, vz2, ...
......
t1, n, xi, yi, zi, vxi, vyi, vzi, ...
......
t2, 1, x1, y1, z1, vx1, vy1, vz1, ...
t2, 2, x2, y2, z2, vx2, vy2, vz2, ...
......
t2, n, xi, yi, zi, vxi, vyi, vzi, ...
......
\end{lstlisting}
}

The output subroutines can be easily integrated into recent versions of {\tt NBODY6} and {\tt NBODY6++}\footnote{\tt http://www.ast.cam.ac.uk/\textasciitilde sverre/web/pages/nbody.htm}, in which option \#46 and \#47 of the input file are reserved for controlling the output file type ({\tt HDF5} or {\tt CSV}) and output frequency, respectively. Detailed instructions for the installation and usage can be found in Appendix~\ref{sec:custom_output}.

\begin{figure}
	\begin{center}
	\includegraphics[scale=0.65]{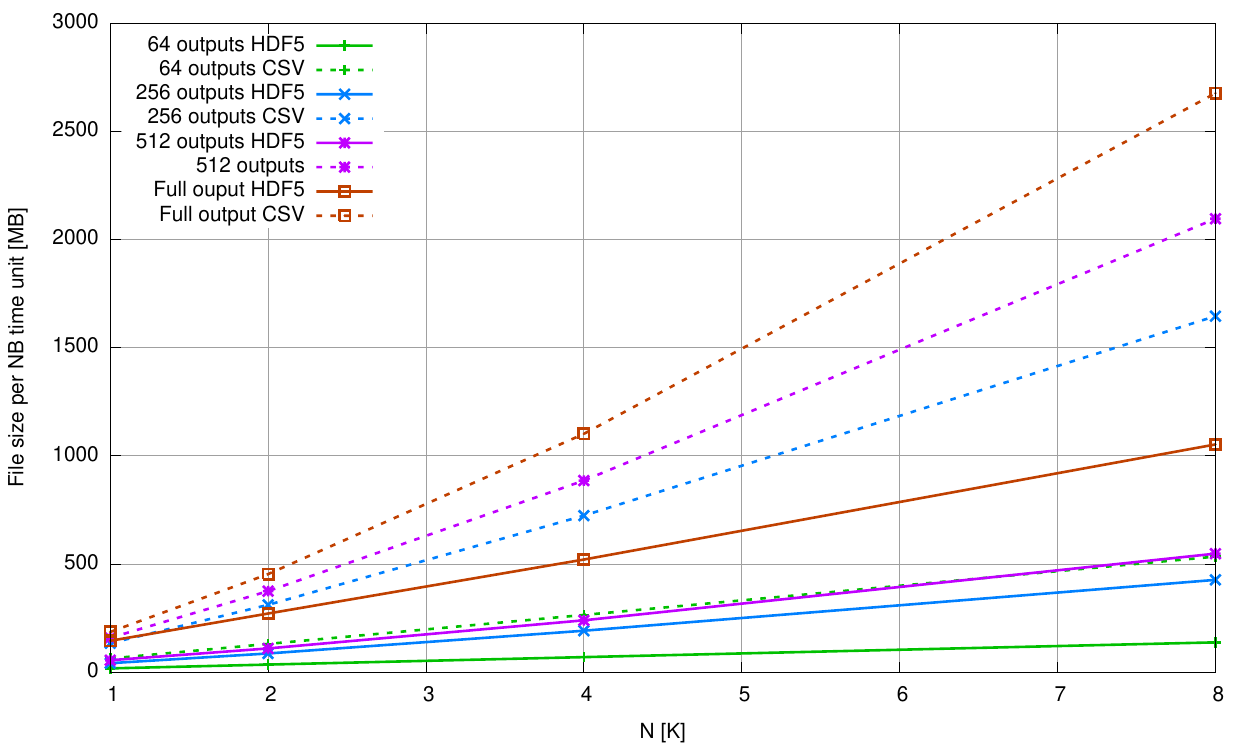}  	
	\caption{The size of output data file for one H\'enon time unit as a function of total particle number and temporal resolution (64, 256, 512 and full output of the BTS data). The solid lines correspond to the output file size of when the output data are written in {\tt HDF5}; the dashed lines correspond to the output file size when the output data are written in {\tt CSV} format. With the increments of particle number and output frequency, the {\tt CSV} file size grows quickly due to its large redundancy of meta data.}
        \label{fig:h5_vs_csv_size}
	\end{center}
\end{figure}

\begin{figure}
	\begin{center}
	\includegraphics[scale=0.65]{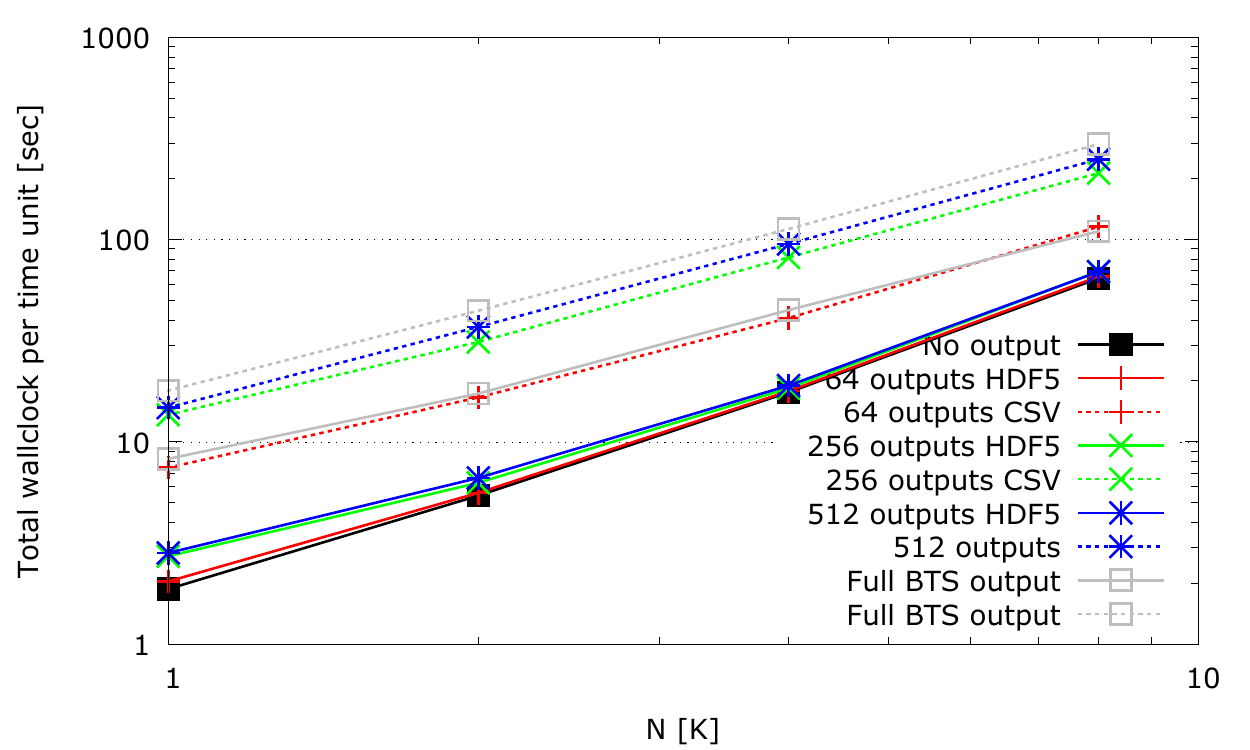}  	
	\caption{Total wall-clock time as a function of total particle number and temporal resolution (64, 256, 512 and full output of the BTS data) for evolving very small systems ($N=1\mathrm{k},2\mathrm{k},4\mathrm{k},8\mathrm{k}$) for one H\'enon time unit. The solid lines correspond to the output file size of when the output data is written in {\tt HDF5}; the dashed lines correspond to the output file size when the output data are written in {\tt CSV} format. This measurement is performed on 4 Intel(R) Core i7-3630QM CPU cores with the {\tt HDF5} library 1.8.11 (without GPU). Even for such small systems, the performance differences of {\tt HDF5} and {\tt CSV} are still well pronounced. The overhead caused by the {\tt HDF5} output routine is negligible, while the corresponding overhead caused by the {\tt CSV} output routine grows quickly.}
    \label{fig:h5_vs_csv_cpu}
	\end{center}
\end{figure}

\begin{figure}
	\begin{center}
	\includegraphics[scale=0.65]{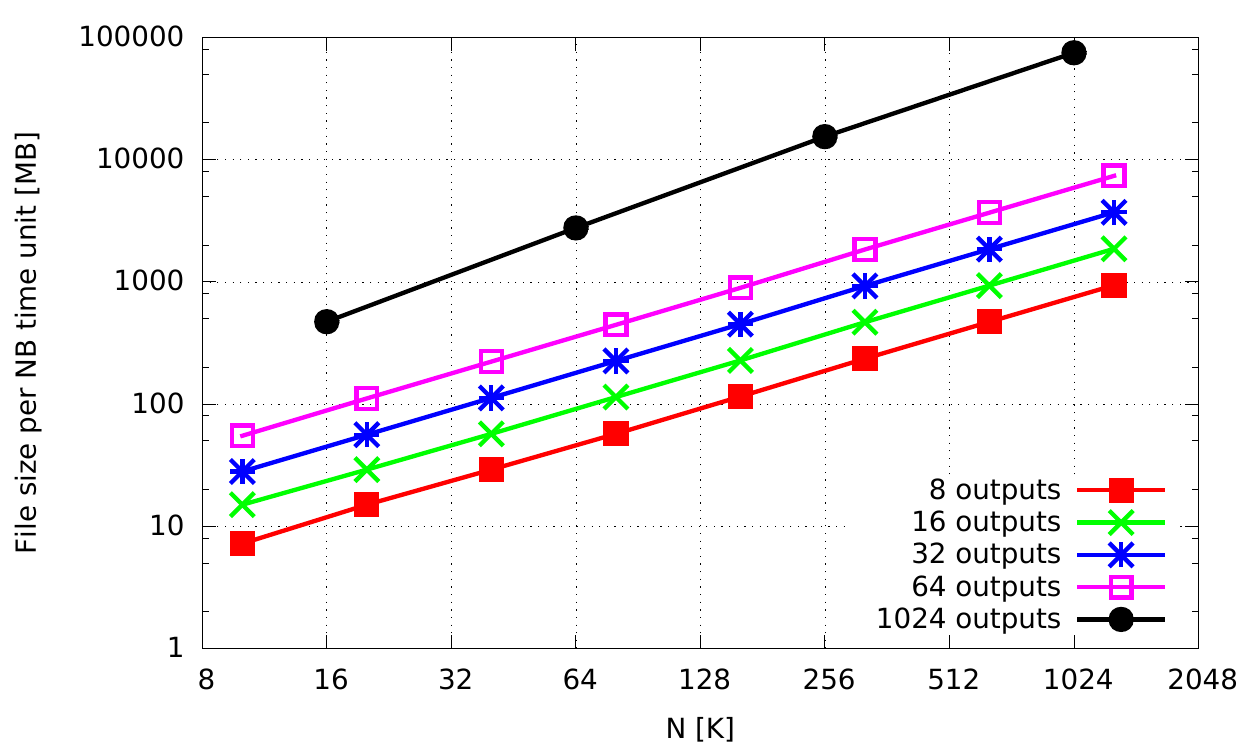}  	
	\caption{The size of output data file by defining temporal resolutions of 8, 16, 32, 64 and 1024 outputs per H\'enon time unit. The output data are written in the HDF5 binary format. This gives a quick estimation of the simulation data size: for example, following up the evolution of a typical $N \sim 100\mathrm{k}$ system for 1000 H\'enon time units with $R_t=3$ (i.e. 8 outputs per H\'enon time unit) corresponds to 100 GB of data size, which can be easily managed, even on personal computers.}
        \label{fig:h5_file_size}
	\end{center}
\end{figure}

The output file sizes of the binary {\tt HDF5} output and text {\tt CSV} output are compared in Fig.~\ref{fig:h5_vs_csv_size}, and the corresponding wall-clock time overheads are shown in Fig.~\ref{fig:h5_vs_csv_cpu}. It is obvious that even for very small systems, the performance difference between {\tt HDF5} files and {\tt CSV} files is well pronounced: the file sizes of {\tt CSV} are generally larger than the corresponding file sizes of {\tt HDF5}, as more data are repeated as meta-data in the {\tt CSV} format. As $N$ increases, they also grow faster than {\tt HDF5}. The overhead of {\tt HDF5} is negligible even for high frequency output, but the overhead of {\tt CSV} is significant. Fig.~\ref{fig:h5_file_size} shows the growth of the file size (a cluster simulation of one H\'enon time unit) as a function of particle number; it also compares the file size dependency on different output frequencies. It shows that at lower output frequencies, the data size grows linearly as a function of $N$, while at high output frequencies (corresponding to $2^{10}=1024$ outputs per h\`enon time unit), the BTS scheme saves a significant fraction of the data rate.

With the scheme described in Section~\ref{sec:temporal}, Fig.~\ref{fig:h5_file_size} shows that the size of output data scales linearly with the total number of particles, thus achieving very high space efficiency and suitable for long-term simulation of very large systems. This scheme may not be able to provide sufficient resolution for highly active particles, as it treats all particles equally. In fact, highly active particles can be well resolved by the scheme described in Section~\ref{sec:spatial}, where the resolution of less active particles is sacrificed.

\section{Applications} \label{sec:applications}
As noted in Section~\ref{sec:ScaleData}, a reasonable tradeoff between output size and loss of information can be achieved by taking the most interesting astrophysical processes into account and then adapting $R_t$ or $R_s$ for data scalings, which makes various applications possible. The resulting datasets can be used for post-simulation processing such as data visualization or data mining; they can also be used to store intermediate simulation data in large scale simulations.

\subsection{Simulation of Planetary Systems in Star Clusters}
The BTS scheme opens new approach for $N$-body simulations involving hierarchical architectures. The stability of planetary systems in star clusters, for example, is of fundamental importance in understand the early stage of planet formation. In fact, star clusters are the building blocks of galaxies \citep{lada2003}. Star formation are believed to be in clusters as giant molecular clouds collapse. The collapse will likely result to circumstellar discs, which are the progenitors of planetary systems. Should planetary systems be formed originally in the star cluster, at least some of them would have been stable enough to survive in the star cluster environment, where the densities are normally much higher than the solar neighbor and close encounters are not rare. The \emph{Kepler} mission has been greatly successful in hunting exoplanets, yet it is worth mentioning that there are only a few exoplanets discovered in star clusters (e.g. Kepler-66, Kepler-67, see \cite{meibom13}). This dichotomy is likely due to the post formation disruptions of planetary systems in star clusters. This problem has been tackled in the previous studies with both direct $N$-body simulations (e.g. \cite{spurzem09}) and Monte-Carlo Simulations (e.g. \citep{Hao13}). Nevertheless, due to the huge range of both dynamical time scales and spatial scales, it is currently only feasible to investigate single planetary systems by treating the star-planet pairs as binaries and employing a regularization technique. Monte-Carlo simulations could indeed extend the study to the multiple planetary system domain, yet the results heavily depend on the quality of close encounter sampling.

With the BTS scheme, it is possible to decouple the dynamics of the whole system into the planetary part and star cluster part, and by which separate the integration of each part. To be specific, the star cluster dynamics can be integrated with a dedicated code such as {\tt NBODY6++}. Having the resulting data from that stored with the BTS scheme, one could then read the stored data, use them to calculate perturbations and plug them into the planetary dynamics code.

\begin{figure}
	\begin{center}
	\includegraphics[scale=0.45]{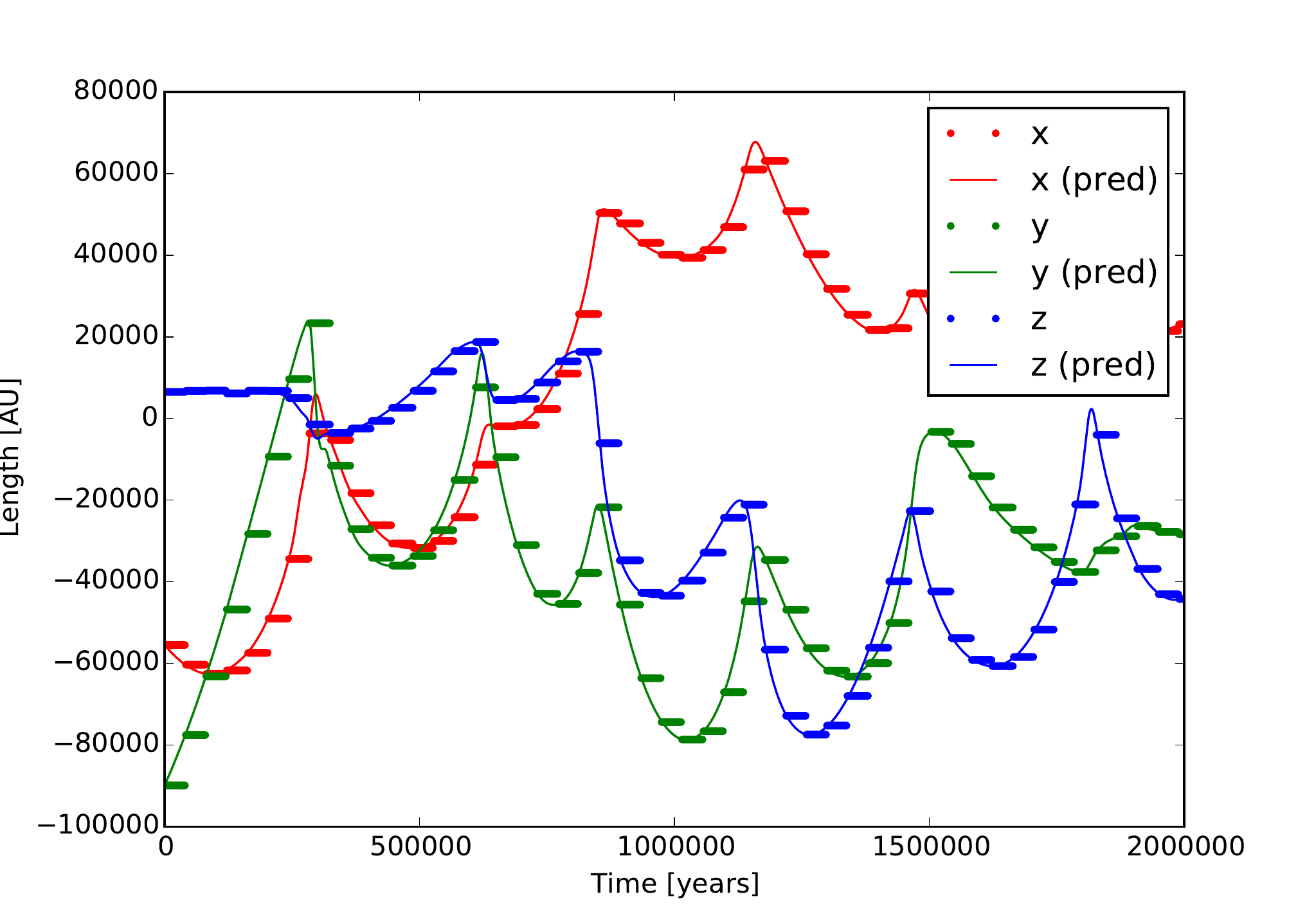}  	
	\caption{Interpolation of the position of a given star. Without interpolation, the position of the star is a step-like function of time (shown with dots). With interpolation, the position is smoothed, allowing velocity kicks to be preciously evaluated (shown with solid curves).}
        \label{fig:interpolation_h5nb6xx}
	\end{center}
\end{figure}

As an example, we implemented the BTS storage scheme with the {\tt HDF5}\footnote{We note especially that {\tt HDF5} is chosen just as an example because it is very prevalent and flexible, other formats such as {\tt SDF} \citep{warren13} exists and can be used in the same way.} file format. The time series data is stored with the {\tt H5Part} scheme \citep{Adelmann08}, as detailed in Table \ref{fig:h5part_file_layout}. The simulations of planetary systems are carried out \emph{after} star cluster simulation is performed and the results are stored in the {\tt HDF5} file. A certain fraction of stars with similar mass are assigned with planetary systems of identical initial configurations. Each planetary system is integrated with {\tt MERCURY6}. As the simulation progresses, the current time $t$ is converted into the the H\'enon time units, and the corresponding step in the {\tt HDF5} file is located thereby. Accelerations at the point where the each planet is located are calculated according to the loaded data, and subsequently be applied as velocity kicks. If $t$ corresponds to the intermediate state between two adjacent time steps, interpolation of $(\mathbf{x}, \mathbf{y}, \mathbf{z})$ will be computed according to equation (\ref{eq:spline0}) to (\ref{eq:spline7}), such that the acceleration at timescales comparable to the typical timescales of planets can be precisely evaluated (as demonstrated in Fig. \ref{fig:interpolation_h5nb6xx}).

Since redundant data is minimized in the BTS storage scheme, we could adopt very high output frequency of star cluster integration data while maintaining reasonable data file size (as shown in Fig. \ref{fig:snapshot_vs_bts}). Together with the septic spline interpolation technique and making full use of all available data in the two adjacent time steps, the velocity kicks can be calculated with very high accuracy and temporal resolution. Furthermore, we parallelize the interpolation on {\tt GPUs} with {\tt Thrust/CUDA}\footnote{\tt http://docs.nvidia.com/cuda/thrust/}. In our simulations, the star cluster has 4000 stars, where  identical planetary systems are assigned to $1\%$ of Solar-type stars. Each planetary system contains the 4 gas giants in the present-day Solar System. The BTS scheme has a temporal resolution $R_t=8$, corresponding to 256 outputs per H\'enon time unit, or roughly $10^4$ years per output. On 2 {\tt Intel Xeon X5650} cores, evolving such a coupled systems for about 1 Myr takes about 12 hours. The code is not fully optimized for the purpose of benchmark, and the actual wall-clock time depends primarily on the frequency of communication between {\tt NBODY6++} and {\tt MERCURY6}. The communication of {\tt NBODY6++} and {\tt MERCURY6} is implemented within the {\tt AMUSE} framework (\citealt{portegies13}; \citealt{portegies09})). The scientific results of this application is presented primarily in Cai et al. (2015, in prep.).

\begin{table*}
\begin{center}
\caption{Internal file layout of the star cluster time series data file}
\begin{tabular}{ccc}
\hline
\hline
Step\# & Attributes (scalar) & Data (vectors)\\
\hline
0 & $t_{0},N_{0},...$ & $\mathbf{x_{0},y_{0},z_{0},x_{0}^{(1)},y_{0}^{(1)},z_{0}^{(1)},x_{0}^{(2)},y_{0}^{(2)},z_{0}^{(2)},x_{0}^{(3)},y_{0}^{(3)},z_{0}^{(3)},m_{0}},...$\\
1 & $t_{1},N_{1},...$ & $\mathbf{x_{1},y_{1},z_{1},x_{1}^{(1)},y_{1}^{(1)},z_{1}^{(1)},x_{1}^{(2)},y_{1}^{(2)},z_{1}^{(2)},x_{1}^{(3)},y_{1}^{(3)},z_{1}^{(3)},m_{1},...}$\\
2 & $t_{2},N_{2},...$ & $\mathbf{x_{2},y_{2},z_{2},x_{2}^{(1)},y_{2}^{(1)},z_{2}^{(1)},x_{2}^{(2)},y_{2}^{(2)},z_{2}^{(2)},x_{2}^{(3)},y_{2}^{(3)},z_{2}^{(3)},m_{2},...}$\\
... & ... & ...\\
$n$ & $t_{n},N_{n},...$ & $\mathbf{x_{n},y_{n},z_{n},x_{n}^{(1)},y_{n}^{(1)},z_{n}^{(1)},x_{n}^{(2)},y_{n}^{(2)},z_{n}^{(2)},x_{n}^{(3)},y_{n}^{(3)},z_{n}^{(3)},m_{n},...}$\tabularnewline
\hline
\end{tabular}
\end{center}
\textbf{Notes.} The time series is organized as {\tt HDF5} groups, and in which vector data and scalar attributes corresponding to a given time step is grouped.
\label{fig:h5part_file_layout}
\end{table*}

\subsection{Long-term Evolution of Massive Globular Cluster with Million Bodies}
Simulations of massive globular cluster in the regime of million bodies and/or million solar masses are made feasible only in recent years, thanks to the exciting evolution of GPU-based high performance computing technology. In a recent research, Wang et al. (2015, in prep.) evolve a globular cluster with $N=1.05\textit{m}$ (950k single stars and 50k binaries) for 12Gyr. With a temporal resolution $R_t = 3$, 8 outputs are generated for each H\'enon time unit, corresponding to 475 MB of data. According to the time scaling of H\'enon time unit to physical time units, the total time of simulation corresponds to about $10^4$ to $10^6$ H\'enon time units, depending on the total mass of the cluster. As such, the total data output of the BTS scheme is roughly 5 TB to 500 TB.

\subsection{Applications for Grid-based Simulations}
BTS-like storage schemes can also be very useful for grid-based simulations. Modern adaptive-mesh-refinement (AMR) codes, such as {\tt Enzo} \citep{bryan14} and {\tt GAMER} \citep{schive10}, adopt the individual time step integration powered by GPU acceleration. The total number of refinement levels is typically around ten, making the evolution time steps of the root level and the highest refinement level differ by a factor of $\sim 1,000$. It is hence impractical to store the entire snapshot at each sub-step.

For example, in the cosmological simulations of wavelike dark matter \citep{schive14}, the dynamical timescale of the solitonic core in each dwarf galaxy is only about 50 Myr. It hence requires $\sim 1.6 \times 10^{4}$ data dumps from redshift one to the present day (assuming 100 dumps per dynamical timescale). Each full snapshot takes about 80 GB in a 1.5 Mpc/h comoving box, with $\sim 10^{10}$ cells in total. The total amount of data in the snapshot scheme thus consume $\sim 1.3$ petabyte. For comparison, if we are mainly interested in the dynamical evolution of one solitonic core, we can utilize the BTS-like storage scheme to only output the core data more frequently. For a solitonic core with a radius of 1 kpc and a simulation resolution of 60 pc, it consumes about 160 kilobyte for one data dump and 2.5 GB in total after redshift one. Accordingly, the storage requirement can be significantly reduced by a factor of $\sim 5 \times 10^{5}$.

\subsection{Data Visualization}
Astronomical data take on a multitude of forms: catalogs, data cubes, images, and simulations \citep{Kent13}. Because of their complexity, they are usually explored using data visualization, which is in fact reorganization of the original data by graphical means. It is particularly useful to illustrate the dynamical evolution of $N$-body systems. Visualization can be done in various ways, from a simple 2D plot to a realistic visual reconstruction of complicated multi-scale astrophysical processes. This simple idea can become challenging in the context of astrophysical data because of the wide dynamical range and large particle number. If the data are stored in a ``compact'' fashion such that only active particles are recorded, as described in Section~\ref{sec:ScaleData}, then the position need to be interpolated using (for example) septic splines as presented in Section~\ref{sec:splines} prior to rendering. Since each particle is interpolated independently, this problem is ``embarrassingly parallel'' and very suitable for GPUs (e.g. programmed in CUDA or OpenCL). It is common that the total number of particles exceeds the total number of pixels on the viewport, and therefore the visualization program should be adjusted to the user's interests. Furthermore, because of the large dynamical ranges, data usually have to be scaled before rendering. For example, the stellar mass $m$ can range from $\sim 0.1\,\mathrm{M}_\odot$ to a much as $\sim 150\,\mathrm{M}_\odot$; the power $P$ emitted by a star is a strong function of its temperature $T$ and radius $R$, as implied by the Stefan-Boltzmann law $P \propto R^2 T^4$.  If $m$ or $P$ are rendered directly on the screen, then massive or bright stars are easily saturated, while light or faint stars are difficult to distinguish.

\begin{figure}
	\begin{center}
	\includegraphics[scale=0.28]{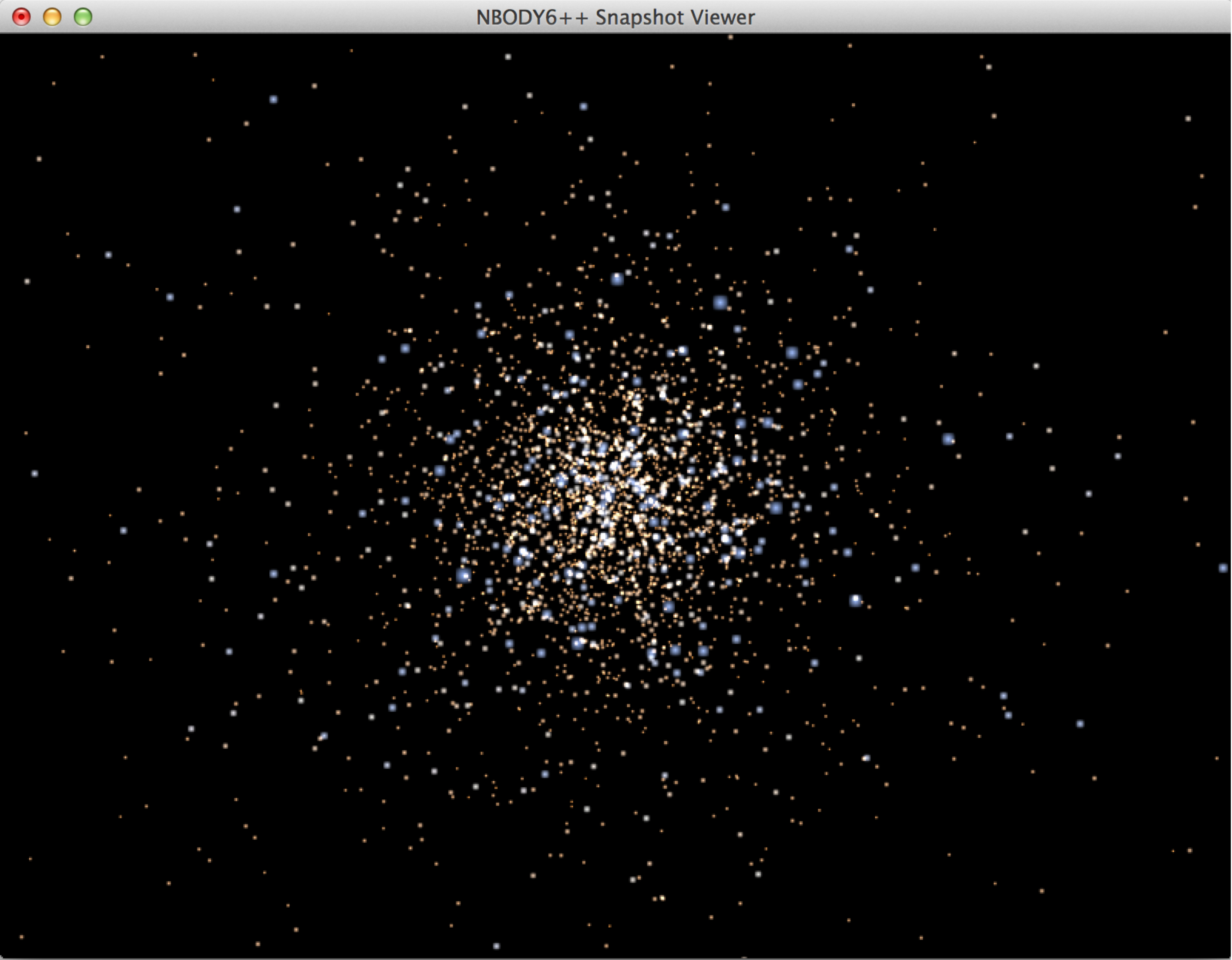}
	\caption{Visualization of {\tt NBODY6++} snapshot with the {\tt vispy} library. The particles are colored according to their temperature, and are sized according to their luminosity.}
        \label{fig:nbody6_viz}
	\end{center}
\end{figure}

\begin{figure}
	\begin{center}
	\includegraphics[scale=0.16]{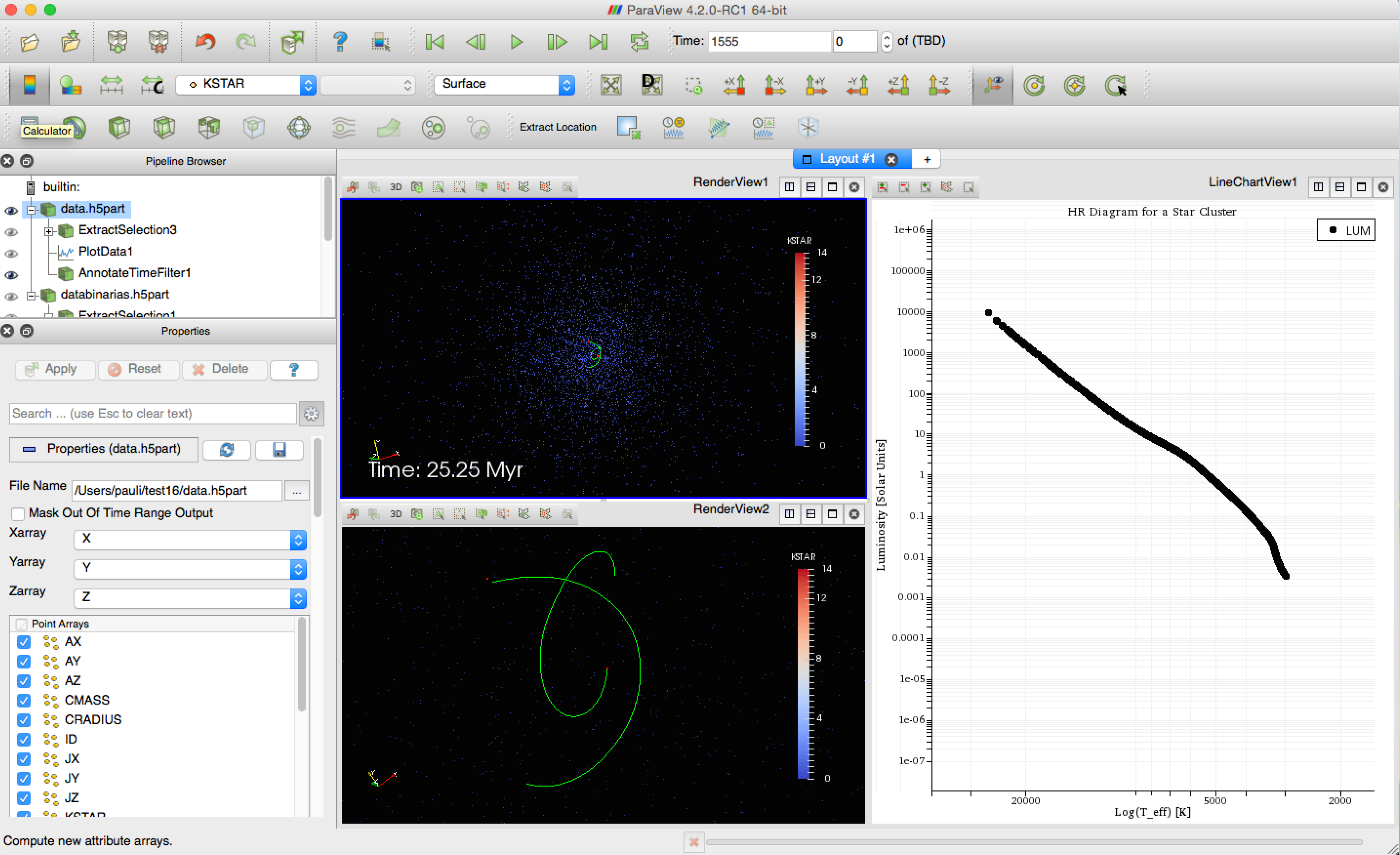}
	\caption{Visualization of {\tt NBODY6++} simulation data with {\tt ParaView}. The figure shows a cluster with $N=5000$ particles (King Model, $W_0=6.0$, Kroupa (2001) initial mass function) The upper left panel shows an overview of the star cluster; the botton-left panel shows the trajectories of the particle of interest (POI), and in this case they are two stellar mass black holes (masses $M_1=10M_{\oplus}$ and $M_2=20M_{\oplus}$). The stars are colored with their stellar types. The corresponding H-R diagram evolves simultaneously with the cluster is shown on the right panel. The data are written in {\tt HDF5} format with {\tt H5Part} scheme, which is supported by {\tt ParaView} via the built-in {\tt H5PartReader}. }
        \label{fig:paraview}
	\end{center}
\end{figure}

In practice, it is usually not enough to recreate the evolution process of an $N$-body system by plotting \emph{only} the coordinates. The velocity vector, mass, size, temperature, luminosity are then expected to be rendered as associated properties of the coordinates, such as color, symbol or size. As an example we adopt the astronomical plotting library {\tt vispy} for the visualization of an {\tt NBODY6++} simulation as Fig.~\ref{fig:nbody6_viz} shows; as another example we also adopt the open source scientific visualization package {\tt ParaView} to visualize the mass spectrum of dense globular cluster simulations, as shown in Fig.~\ref{fig:paraview}.

\section{Conclusion}
We present the Block Time Step (BTS) storage scheme for the data management of astrophysical $N$-body simulations, inspired by the individual time step integration scheme \citep{makino88}. This is an urgent response to the ever increasing challenges posed by modern highly computationally expensive simulations. By adopting the BTS storage scheme, the growth of data can be dramatically scaled down from $N^2$ to $N^{4/3}$ (for the Plummer model). Depending on the usages of simulation, it is not necessary for all integration data to be recorded. Instead, a resolution parameter can be defined either in the space domain or in the time domain, which offers the flexibility of data scaling. Apart from theoretical analysis and predictions, dedicated simulations are carried out and the results are consistent with the theory. Our I/O performance benchmark of the binary {\tt HDF5} and the \textsc{ascii} {\tt CSV} format shows that binary formats are generally more preferable to store large and complicated datasets, yet for lightweight datasets text files exhibit their convenience for data analysis and portability. A list of astrophysical quantities for particles with potential user interests is proposed, and some of these quantities are visualized with open-source packages and libraries such as {\tt ParaView}, {\tt GLnemo2} and {\tt s2plot}.

The growth of numerical simulation scales and data rates implies that not only computations, but also data storage, visualization and analysis need to be carried out distributively. Open source packages currently provide strong support for the technical implementation of these distributed systems, yet, to implement them into astrophysics-driven systems, adaptations need to be made according to the specific astrophysical context. This paper therefore addresses the concerns and possible solutions. Typical applications of the proposed scheme on astrophysical scenarios such as simulations of planetary systems in star cluster, cosmological grid-based simulations, long-term evolution of million solar masses globular clusters and scientific visualizations are presented as well.

Our discussion is primarily focused on particle-based direct $N$-body simulations.  The philosophy behind the proposed scheme is to ``\emph{apply proper scaling to the simulation data to provide fine-grain control of the resolution of scientifically interesting data while suppressed the uninteresting ones}''. Despite the different algorithms used in other kinds of astrophysical simulations, such as hydrodynamics simulations, tree codes, adaptive mesh refinement (AMR) codes, Monte-Carlo simulations, and many other new algorithms under development, the methodology addressed in this paper is transferable to a wide range of scenarios.

\section*{Acknowledgments}
We wish to thank the anonymous referee for her/his constructive comments that helped to improve the manuscript considerably.

We acknowledge support by NAOC CAS through the Silk Road Project and (RS)
through the Chinese Academy of Sciences Visiting Professorship for Senior
International Scientists, Grant Number $2009S1-5$.
The special GPU accelerated supercomputer {\tt laohu} at the Center of Information
and Computing at National Astronomical Observatories, Chinese Academy of
Sciences, funded by Ministry of Finance of People's Republic of China under the
grant $ZDYZ2008-2$, has been used for some of the largest simulations.

We are grateful for support by Sonderforschungsbereich SFB 881 "The Milky Way System"
of the German Research Foundation (DFG),
through subproject Z2 and the GPU cluster Milky Way at FZ J\"ulich, and for the
support of the visit of MXC in Heidelberg.

We thank Hsi-Yu Schive for the information of cosmological grid-based simulation. We thank Peter Berczik, Long Wang, Sverre Aarseth, Marcel Zemp and Siyi Huang for useful discussions.

MBNK was supported by the Peter and Patricia Gruber Foundation through the PPGF fellowship, by the Peking University One Hundred Talent Fund (985), and by the National Natural Science Foundation of China (grants 11010237, 11050110414, 11173004). This publication was made possible through the support of a grant from the John Templeton Foundation and National Astronomical Observatories of Chinese Academy of Sciences. The opinions expressed in this publication are those of the author(s) do not necessarily reflect the views of the John Templeton Foundation or National Astronomical Observatories of Chinese Academy of Sciences. The funds from John Templeton Foundation were awarded in a grant to The University of Chicago which also managed the program in conjunction with National Astronomical Observatories, Chinese Academy of Sciences.

PA acknowledges financial support of FONDECYT 3130623 and CAS Visiting Fellowship for researchers from developing countries.
PA was also funded by Chinese Academy of Sciences President's International Fellowship Initiative; Grant No. \textit{2014FFJB0018}.

\appendix

\section{Custom Output Subroutines for {\tt NBODY6}} \label{sec:custom_output}
We implemented the {\tt HDF5} and {\tt CSV} custom data format output subroutines for {\tt NBODY6} and {\tt NBODY6++}, both for benchmark purpose and the need of long term data management. The subroutines can be downloaded from {\tt http://silkroad.bao.ac.cn/\textasciitilde maxwell/hdf5}. The integration is trivial: (1) compile and install the {\tt HDF5} library from the source code, which can be obtained from {\tt http://www.hdfgroup.org/HDF5/release/obtainsrc.html}; (2) copy the custom output subroutines source code {\tt custom\_output.f} to the {\tt Ncode} directory of {\tt NBODY6}; (3) modify the Makefile to add the {\tt custom\_output.f} file into the list of source files; (4) call the subroutine by adding one line into the {\tt intgrt.f} (or {\tt intgrt.omp.f} for the {\tt GPU2} version). (5) Add a common block to the ``hrplot.f'' so that stellar evolution data can also be dumped to the output. More detailed instruction can be found in the {\tt README} file of of the downloaded package. In the {\tt NBODY6} input file, option \#46 and \#47 are used to control the output file type, respectively, as shown in Table~\ref{table:custom_output}.

\begin{table}
\begin{center}
\caption{Fine-grain control of output frequency and file format for {\tt NBODY6}.}
\begin{tabular}{ccc}
\hline
\hline
Option & Meaning  \\
\hline
{\tt KZ(46)=1} & Output BTS data as {\tt HDF5} (active particle only) \\
{\tt KZ(46)=3} & Output BTS data as {\tt HDF5} (all particles)  \\
{\tt KZ(46)=2} & Output BTS data as {\tt HDF5} (active particle only) \\
{\tt KZ(46)=4} & Output BTS data as {\tt CSV} (all particles)  \\
{\tt KZ(47)=$R_t$} & The output frequency is $2^{R_t}$ times per H\'enon time unit \\
\hline
\end{tabular}
\end{center}
\label{table:custom_output}
\end{table}

\section{Visualization techniques of the {\tt HDF5}-based BTS data}
The {\tt HDF5} data generated by the custom output subroutines described in Appendix~\ref{sec:custom_output} can be visualized \emph{directly} with {\tt ParaView}. The {\tt H5Part} reader is included in the {\tt ParaView} 4.x distribution, but it is not activated by default. To enable it, users may navigate to the main menu and click ``Tools | Manage Plugins'', and then find the {\tt H5PartReader} and select ``Auto Load''. After that one will be able to load the {\tt HDF5} simulation datasets from the {\tt Open} menu. After loading the file, users may select the X, Y and Z arrays from the drop-down list for visualization.

We also implemented a {\tt vispy}-based visualization script for the simulation datasets, which can be downloaded from {\tt http://silkroad.bao.ac.cn/\textasciitilde maxwell/hdf5}.
\bigskip

\phantom{a}

\end{document}